\newif\ifdouble
\doublefalse

\ifdouble
  \documentclass[journal,twocolumn,10pt]{IEEEtran}
\else
  \documentclass[draftcls,onecolumn,12pt]{IEEEtran}
\fi

\newif\ifsmallalg
\smallalgfalse

\usepackage[cmex10]{amsmath}
\usepackage{amssymb}
\usepackage[latin1]{inputenc}
\usepackage{color}
\usepackage{cite}
\usepackage{subfigure}

\usepackage[algoruled,linesnumbered]{algorithm2e}

\usepackage[dvips]{graphicx}

\newcommand{\vva}{``}
\newcommand{\mc}{\mathcal}
\newcommand{\beq}{\begin{equation}}
\newcommand{\eeq}{\end{equation}}

\newcommand{\Pfa}{\mathsf{P_{fa}}}
\newcommand{\Pd}{\mathsf{P_d}}

\newcommand{\wi}{0.45\textwidth}
\newcommand{\mb}{\mathbf}
\newcommand{\bm}{\boldsymbol}
\newcommand{\pp}[1]{_{\scriptscriptstyle{(#1)}}}
\newcommand{\ap}[1]{^{(#1)}}

\newcommand{\maggmin}{ \mathop{\stackrel{\mc H_1}{\gtrless}}_{\mc H_0}  }

\newcommand{\AC}{\mathtt{AC}}
\newcommand{\yr}[1]{\mathbf{y}[#1]}  
\newcommand{\zr}[2]{\mathbf{z}_{#1}[#2]}
\newcommand{\ar}[1]{\mathbf{a}[{#1}]}  %{\mathbf{Y}_{[#1,\cdot]}}
\newcommand{\yc}[1]{\mathbf{y}({#1})}  %{\mathbf{Y}_{[\cdot,#1]}}
\newcommand{\zc}[2]{\mathbf{z}_{#1}({#2})}
\newcommand{\ac}[1]{\mathbf{a}({#1})}  %{\mathbf{Y}_{[\cdot,#1]}}
\newcommand{\diag} {\mathrm{diag}}
\newcommand{\Diag} {\mathrm{Diag}}
\newcommand{\ewtimes} {\odot}
\newcommand{\ewfrac} {\oslash}
\newcommand{\rnorm} {\mathtt{RN}}

\newcommand{\pma}{ ^\mathsf{PM1}}
\renewcommand{\pmb}{^\mathsf{PM2}}
\newcommand{\pmc}{^\mathsf{PM3}}
\newcommand{\laa}{ ^\mathsf{LA1}}
\newcommand{\lab}{^\mathsf{LA2}} 
\newcommand{\enu}{e_{\mathsf{num}}}
\newcommand{\ede}{e_{\mathsf{den}}}
\newcommand{\ide}{^{\mathsf{id}}}
\renewcommand{\th}{\vartheta}

\newcommand{\TR}{T_{\mathsf {R}}}
\newcommand{\TG}{T_{\mathsf {G}}}
\newcommand{\TS}{T_{\mathsf {S}}}
\newcommand{\TJ}{T_{\mathsf {J}}}

\newtheorem{proposition}{Proposition}

\begin{document}

%\title{Decentralized Largest Eigenvalue Test for Multi-Sensor Signal Detection}
%\title{Decentralized Eigenvalue-based Signal Detection via Average Consensus} % keep for journal paper
\title{Decentralized Eigenvalue Algorithms for Distributed Signal Detection in Cognitive Networks} 
%\title{Decentralized Eigenvalue Algorithms in Wireless Networks} 
%\title{Decentralized Computation of the Eigenvalues of Multi-Sensor Covariance Matrices}

%
%
\author{
{Federico Penna and S{\l}awomir Sta\'{n}czak
\thanks{The authors are with the Fraunhofer Institute for Telecommunications, Heinrich-Hertz-Institute,
Einsteinufer 37, 10587 Berlin, Germany.
 Email:  \{federico.penna, slawomir.stanczak\}@hhi.fraunhofer.de. The work was supported by the German Research Foundation (DFG) under grant STA864/3-2 and by the German Federal Ministry of
Education and Research under grant 01BU1224.
 Part of the contents of this paper were presented at the IEEE Globecom 2012 conference \cite{gc12} and the IEEE Dyspan 2012 conference (poster session) \cite{dyspan12}.
 } 
 }
}

%\markboth{Submitted to ....}%
%{Submitted to ....}

\maketitle

\thispagestyle{empty}

\begin{abstract}
  In this paper we derive and analyze two algorithms -- referred to as decentralized power method (DPM) and decentralized Lanczos algorithm (DLA) -- for distributed computation of one (the largest) or multiple eigenvalues of a sample covariance matrix over a wireless network.
The proposed algorithms, based on sequential average consensus steps for computations of matrix-vector products and inner vector products, are first shown to be equivalent to their centralized counterparts in the case of exact distributed consensus. Then,  closed-form expressions of the error introduced by non-ideal consensus are derived for both algorithms. The error of the DPM is shown to vanish asymptotically under given conditions on the sequence of consensus errors. 
Finally, we consider applications to spectrum sensing in cognitive radio networks, and we show that virtually all eigenvalue-based tests proposed in the literature can be implemented in a distributed setting using either the DPM or the DLA.
Simulation results are presented that validate the effectiveness of the proposed algorithms in conditions of practical interest (large-scale networks, small number of samples, and limited number of iterations).
\end{abstract}


\section{Motivation and Related Work}
\label{intro}

Computing the eigenvalues of sample covariance matrices is a fundamental problem in signal processing, with  applications including multi-sensor spectrum sensing in cognitive radio networks (see Section \ref{app}).
%In general, given a (possibly large-dimensional) dataset composed of signal samples received at multiple antennas or sensor nodes, the eigenvalues of the sample covariance matrix can be used for statistical inference of the number and power of the signal components as well as the noise level. 
%Such information makes it possible to identify the presence of signal sources in cognitive radio scenarios,
% or abnormalities in the traffic distribution in computer network analysis.
 Given the increasing popularity of dense, large-scale wireless sensor networks, applications of eigenvalue-based inference techniques in distributed settings are of great interest. However, most of the eigenvalue-based techniques proposed in the existing literature assume a centralized architecture, where the samples received by different nodes are forwarded to a fusion center that is in charge of constructing the sample covariance matrix and computing the relevant test statistics.
 This traditional architecture  has several drawbacks: 
 it requires a fusion center with high computational capabilities, therefore it does not support applications in which 
 different nodes may be chosen as fusion centers at different times;
 relying on one node only, it is vulnerable to hardware failures, Byzantine faults, or  attacks from malicious users;
it is not efficient in case of multi-hop networks, where some nodes may be many hops away from the fusion center; 
and it lacks scalability, because a growing number of nodes in the network may result in congestion of the communication channel with the fusion center.

For these reasons, we  seek a  decentralized method to compute the eigenvalues of sample covariance matrices over a wireless network, such that
  the computational effort is distributed across multiple nodes
  and the many-to-one communication protocol is replaced by a more scalable neighbor-to-neighbor protocol.
    In this paper we propose two solutions  based on decentralized implementations of iterative eigenvalue algorithms --  the {power method} (PM, see Section \ref{sec_pm}) and the Lanczos algorithm (LA, see Section \ref{sec_la}).  
    Decentralized versions of such algorithms are obtained by applying average consensus (AC, see Section \ref{sec_ac}) as a subroutine to perform those computations that involve combining the data of different nodes.
    Once local estimates of the eigenvalues of interest are computed at every node, statistical tests for signal detection can be performed locally by each node.

  The contribution of this paper is related to that of \cite{scaglione_ieee}, where a decentralized algorithm based on the Oja-Karhunen recursion is proposed to track the eigenvectors of a covariance matrix. 
  Our work adopts the same idea of computing inner vector products  through AC (and further extends this approach to  matrix-vector products), but  differs from \cite{scaglione_ieee} in the following sense:
    \textit{(i)} we compute {eigenvalues} instead of (possibly, in addition to) eigenvectors, thus adapting the well-established theory of eigenvalue-based detection to decentralized network settings;
  \textit{(ii)} we focus on {detection} (decision based on $N$ received samples per sensor) instead of sequential tracking (update of eigenvector estimates at every new sample). This makes our approach suitable for spectrum sensing and other signal detection applications. 
  A similar methodology for distributed matrix multiplication via AC has been recently  used also in 
 \cite{zazo}, where a decentralized expectation-maximization algorithm is derived for static linear Gaussian models. 
 
 Other recent related works are \cite{dpca} and \cite{gpca}, both proposing decentralized implementations of principal component analysis (PCA) over wireless networks. The first approach \cite{dpca}  relies on the assumption of decomposable Gaussian graphical models, i.e., it requires data sample to be \textit{(i)}  multivariate Gaussian distributed, and \textit{(ii)} decomposable into two or more conditionally independent ``cliques''. The global eigenvalue decomposition (EVD) problem is thus broken down into a sequence of local (clique-wise) EVD subproblems.
The second approach \cite{gpca} combines the power method with the concept of sparsification to achieve an efficient distributed computation of the eigenvectors and eigenvalues of a symmetric matrix. However, this approach is based on the assumptions that  \textit{(i)} each node has access to a full row of the matrix (which is not the case with the sample covariance matrix considered in our model, see Section \ref{ps}), and  \textit{(ii)} the network graph is completely connected (i.e., a direct link exists between any pair of nodes).
Our approach, on the contrary, does not require any of the aforementioned assumptions.

    The paper is organized as follows.
  A formal statement of the problem is provided in Section \ref{ps}. 
  Section \ref{prelim} contains mathematical preliminaries about the algorithms used in this work (AC, PM, and LA).
  We then introduce the proposed algorithms in Section \ref{algo} and analyze their performance and complexity in Section \ref{perf}.
  We finally discuss two practical applications of the proposed algorithms in Section \ref{app} and present numerical results in Section \ref{simul}. 
  Concluding remarks are provided in Section \ref{concl}.
  
%%  
%%  
%  The rest of the paper is organized as follows.
%  The problem and the mathematical model are presented in Section \ref{ps}.
%  The proposed algorithms are described in Section \ref{algo} and evaluated by simulations in Section \ref{sim}.
%  Concluding remarks are provided in Section \ref{concl}.
  
% In a previous work, a decentralized implementation of multi-sensor energy detection was proposed, exploiting the average consensus algorithm}. Average consensus is perfectly suited for cooperative energy detection, as the relevant test statistic is exactly the average of the energies measured at different sensors. The problem becomes more complex in the case of eigenvalue-based detection, because computation of a covariance matrix or of its eigenvalues can not be immediately reduced to an averaging problem.
% Nevertheless, in this paper we resort to iterative eigenvalue algorithms -- the power method and the Lanczos algorithm -- and we reduce them to 
% In this way, we obtain two  algorithms that compute iteratively, in a fully decentralized fashion, the largest eigenvalue of the received signal covariance matrix, i.e., the test statistic needed to perform the Roy's largest root test  ...
% 
% The key computational step of our algorithms is the computation of inner products via average consensus. Such idea is  
% inspired by \cite{scaglione_ieee}, where a decentralized algorithm was proposed for subspace tracking (i.e., tracking the eigenvectors of a covariance matrix). 
% Our work uses a similar approach 

\section{Problem Statement}
\label{ps}

Consider a wireless network consisting of $K$ sensor nodes. During a given time interval (sensing period), each node collects $N$  complex  signal samples.
The global received sample matrix is denoted by
\beq
\label{Y}
\mb Y = [\yc{1}, \hdots, \yc{N}] = \left[ 
\begin{array}{c}
\yr{1}^T\\
\vdots \\
\yr{K}^T
\end{array}
\right] \in \mathbb C^{K\times N} ,
\eeq 
where symbols $\yc{\cdot} \in \mathbb C^{K}$ and $\yr{\cdot} \in \mathbb C^{N}$
%\beq
%\yc{n} = [y_1(n), \hdots, y_K(n)]^T
%\eeq
%and
%\beq
%\yr{k} = [y_k(1), \hdots, y_k(N)]^T
%\eeq
are used to denote, respectively, the columns and (transpose) rows of $\mb Y$.
Physically, column $\yc{n}$ contains the samples received by all nodes at time $n$, while row  $\yr{k}^T$ contains all samples available at node $k$ at the end of the sensing period.
We then define the sample covariance matrix as
\beq
\label{R}
\mb R \triangleq   \frac 1 N \mb Y\mb Y^H.
\eeq
Let $\lambda_1 \geq \hdots \geq \lambda_K \geq 0$ be the eigenvalues of $\mb R$, without loss of generality sorted in decreasing order, and $\mb u_1, \hdots, \mb u_K$ the corresponding eigenvectors.
The problem addressed in this work can be stated as follows:
how can a network compute (or estimate) one or more of the above eigenvalues without a fusion center that collects all samples $\mb Y$, and without explicitly constructing the sample covariance matrix $\mb R$?
Before presenting the proposed solution, we introduce some preliminary definitions and basic concepts about distributed consensus, power method, and Lanczos algorithm.
The notation used throughout the paper is summarized in Table \ref{notation}.

\begin{table*}[tbp]
\centering
 \small 
\begin{tabular}{ll}
\hline
\textbf{Symbol} & \textbf{Definition} \\ 
\hline
\hline
$\| \mb v \|$ &  Euclidean ($\ell^2$) norm of  vector $\mb v$ \\ 
\hline 
$^T$, $^*$, $^H$ & Transpose, complex conjugate, and Hermitian operators \\
\hline
$\ac{n}$, $\ar{k}^T$ & Respectively, column $n$ and row $k$ of matrix $\mb A$\\
\hline
$\ewtimes$, $\ewfrac$ %, $\ewpow{a}$ 
& Element-wise vector multiplication and division\\%, and $a$-power \\
\hline
$\mb I_m$, $\mb 1_m$ & Identity matrix and column vector of ones of size $m$ (subscript is omitted when not ambiguous) \\
\hline
$\diag(\mb A)$ & Column vector with diagonal elements of a square matrix $\mb A$: i.e., $\diag(\mb A) = (\mb A \ewtimes \mb I) \cdot \mb 1 $\\
\hline
$\Diag(\mb v)$ & Square matrix with vector $\mb v$ as main diagonal: i.e., $\Diag(\mb v) = (\mb v \cdot \mb 1^T) \ewtimes \mb I $\\
\hline
$\rnorm(\mb A)$ & Vector of $\ell^2$ square norms of the rows of $\mb A$: i.e., $\rnorm(\mb A) \triangleq \left[ \|\ar{1}\|^2, \hdots, \|\ar{K}\|^2 \right]^T = (\mb A^* \ewtimes \mb A ) \cdot \mb 1$ \\
\hline
$\AC_m^t(\cdot)$ & AC function with global input/output for all nodes (Eq. \ref{acdef})\\ % $\mb Z_0$, $\mb Z_t \in \mathbb C^{K \times m}$ (resp., $\mb z_0$, $\mb z_t \in \mathbb C^{K}$ if $m=1$)\\
							& $m$ = input vector size, $t$ = AC iterations or running time (no superscript = ideal AC)  \\
\hline
$\AC_m^t[k](\cdot)$ & AC function with local input/output for node $k$ (Eq. \ref{ack}) \\
 %, i.e., input/output = $k$-th row of input/output of $\AC_m^t(\cdot)$\\
\hline
\end{tabular}
\normalsize
\caption{Notation.}
\label{notation}
% \vspace{-1cm}
\end{table*}

\section{Preliminaries}
\label{prelim}

\subsection{Average Consensus}
\label{sec_ac}

Assume that the  network nodes and their links form a connected undirected graph.
Under such an assumption, it is possible to define a distributed AC algorithm over the network. By distributed AC we mean any algorithm whose output for all nodes converges to the average of the initial values of the individual nodes. 
A large variety of AC algorithms have been proposed in the literature (see \cite{ac_overview} for a survey),
both with synchronous \cite{boyd,boyd2} and asynchronous protocols \cite{murray,gossip,consprop}. Extensions to noisy message exchange and link failures include \cite{kar,nedic,tahbaz,barbarossa,garin}, and methods for consensus acceleration have been proposed in \cite{renato_tsp2010,renato_cheb,comac_meng}. 
It is worth noting that, under the assumption of fixed network topology and noiseless communication, exact consensus can be achieved in a finite number of steps: see for example \cite{finitetime}.

In this paper we do not adopt a specific AC algorithm, but rather  take a general approach. We model
 the result of a generic AC routine by a function $\AC_m^t: \mathbb C^{K \times m} \to \mathbb C^{K \times m}$, where $m$ is the size of the input vectors at each node\footnote{An AC algorithm with vector inputs involves exchanging $m$ scalar numbers at every iteration, and returns the average of the input vectors over all network nodes. As such, it involves the same number of messages as scalar AC, but with a larger payload.} 
 and $t$ is the number of iterations or the averaging time.\footnote{For sake of generality, we do not specify whether the adopted AC routine should be synchronous or asynchronous. In the first case $t$ is a discrete  number of iterations; in the second case it is the running time of the algorithm, that is related probabilistically to the number to clock ticks (see for example \cite{gossip}).} 
 Denote the global input (or initial value) matrix 
by 
\beq
\mb Z_0 = [\zc{0}{1}, \hdots, \zc{0}{m}] = \left[ 
\begin{array}{c}
\zr{0}{1}^T\\
\vdots \\
\zr{0}{K}^T
\end{array}
\right]
\in \mathbb C^{K \times m},
\eeq
defined in analogy with (\ref{Y}), such that the $k$-th row represents the samples available at node $k$. 
Then, the output of the AC function at time $t$ is defined by
\beq
\label{acdef}
\mb Z_t = \AC_m^t (\mb Z_0) \triangleq \frac 1 K \mb 1 \mb 1^T \mb Z_0 + \mb E_t,
\eeq 
where $\mb 1$ is a column vector of ones of size $K$, and 
\beq
\mb E_t = [\mb e_t(1), \hdots, \mb e_t(m)]  \in \mathbb C^{K \times m}
\eeq 
is an error term depending on the AC time $t$ and on the specific AC method. In general, $\mb E_t$ can be assumed to be either bounded (at least statistically, see \cite{murray,gossip,consprop,boyd2,tahbaz,garin,nedic,renato_cheb}) or equal to zero (in the case of finite-time AC \cite{finitetime}).

%If finite-time AC algorithms are used, assuming fixed topology and no noise (see \cite{finitetime}, \cite[Lemma 1]{renato_tsp2010}), then there exists a $t \leq K$ such that $\mb E_t = \mb 0$. 
%For other types of AC algorithms that provide asymptotic convergence, bounds on the first and second moment of $\| \mb e_t(n) \|$ ($1\leq n \leq m$) for finite $t$ exist, e.g., 
%\beq
%\mathbb E [ \| \mb e_t(n) \|^\nu ] \leq \delta_{\nu,t} ,
%\eeq
%where $\nu=1$ or $\nu=2$. The bounds $\delta_{\nu,t}$ depend on the adopted AC algorithm and on factors such as  the network topology, the number of iterations or running time $t$, and   the norm of the  initial vector  $\| \zc{0}{n} \|$.
%For practical examples, see \cite{murray,gossip,consprop,boyd2,tahbaz,garin,nedic,renato_cheb}, etc. 
%In this paper we assume that, if asymptotic AC algorithms are used,  the value of $t$ is chosen so as to guarantee a required precision of the AC output.

In the following, it is sometimes convenient to express the input and the output of AC for a single node $k$. 
For this purpose, we define a function $\AC_m^t[k]: \mathbb C^{1 \times m} \to \mathbb C^{1 \times m}$
having as input/output the $k$-th rows of the input/output matrices of the global function $\AC_m^t(\cdot)$ defined in (\ref{acdef}). Thus, we can write
\beq
\label{ack}
\zr{t}{k}^T = \AC_m^t[k] (\zr{0}{k}^T).
\eeq
%where  is a function having as input/output the $k$-th rows of the input/output matrices of the global function $\AC_m^t(\cdot)$ defined in (\ref{acdef}).
 % matrices $\mb Z_t$ and $\mb Z_0$, respectively.
When AC is applied to scalar arguments ($m=1$), the global input-output relation is written as $\mb z_t = \AC_1^t(\mb z_0)$, with  $\mb z_0, \mb z_t \in \mathbb C^K$.
Finally, if $\mb E_t=\mb 0$  in (\ref{acdef}), we call the AC routine ``ideal'' and we use the notation
 $\AC_m(\cdot)$ (without superscript).

\subsection{Power Method}
\label{sec_pm}

The PM is a well-known iterative algorithm that, given a square matrix,  converges to the eigenvector associated with the largest eigenvalue of the matrix \cite{golub}.
The iteration, applied to the sample covariance matrix $\mb R$, can be written as
\beq
\label{pmiter}
\mb v \pp{j+1} =  \mb R \mb v\pp{j},
\eeq
where $\mb v\pp 0 \in \mathbb C^{K}$ is an arbitrary starting vector.
By (\ref{pmiter}),  the $j$-th iteration can be written as $\mb v \pp{j} = \mb R^j \mb v\pp 0$ and, for $j \to \infty$, the vector $\mb v \pp j$ converges to a multiple of the eigenvector $\mb u_1$.
The convergence rate of the algorithm is $O(({\lambda_2}/{\lambda_1})^M)$ \cite{golub}.
After $M$ iterations, the largest eigenvalue of $\mb R$ can be approximated by
\beq
\label{hatl1}
\hat \lambda_1 =   \dfrac{ \mb v\pp M^H \mb R \mb v\pp{M}} {\mb v\pp M^H \mb v\pp{M}}.
\eeq

%, applied to the sample covariance matrix $\mb R$, is reported in Alg. \ref{powermethod}.
%Superscripts in brackets represent the iteration index (e.g.,  $\mb v\pp j$), and $\hat \lambda_1$ denotes an estimate of the eigenvalue $\lambda_1$.

% \subsubsection*{Remarks on notation}
% In the following,
%superscripts in brackets represent iteration indices;
%$\lambda_i\pp j$  denotes an estimate of $\lambda_i$ (the $i$-th largest eigenvalue of $\mb R$) obtained at iteration $j$; 
%$\lambda_{i,k}\pp j$ is the decentralized version of $\lambda_i\pp j$ at node $k$;
%$\mb u_i\pp j$ represents an estimate of $\mb u_i$ (the eigenvector associated to $\lambda_i$) at iteration $j$;
%$\mb {\dot u}_i\pp j$ is an unnormalized version of  the unit-norm vector $\mb u_i\pp j$;
%$ u_{i,k}\pp j$ is the $k$-th element of vector $\mb u_i\pp j$;
%similarly, for any generic vector $\mb v$, the $k$-th element is denoted by $v_k$.
%In decentralized algorithms, all quantities marked by subscript $k$ are computed locally by node $k$. 

%\begin{algorithm}[h!]
%% \ifsmallalg \small \fi
%	\caption{Power method}
%	\label{powermethod}
%	\SetKwInOut{Input}{Input}\SetKwInOut{Output}{Output}
%	\Input{Matrix $\mb R \in \mathbb C^{K\times K}$; number of iterations $M$; unit-norm starting vector $\mb v\pp{0}$}
%	\Output{Largest eigenvalue estimate $\hat\lambda_1$}
%			\BlankLine
%		\For{iteration $j=1$ to $M$} {
%		  $\mb v \pp{j} \la  \mb R \mb v\pp{j-1}$\;  \label{pmu}
%}
%  $  \hat \lambda_1 \la   \dfrac{ \mb v\pp M^H \mb R \mb v\pp{M}} {\mb v\pp M^H \mb v\pp{M}} $\; \label{pml}
% 		\normalsize
%\end{algorithm}    

\subsection{Lanczos Algorithm}
\label{sec_la}

%
%\begin{algorithm}[tb]
%%\ifsmallalg \small \fi
%	\caption{Original Lanczos algorithm}
%	\label{lanczos}
%	\SetKwInOut{Input}{Input}\SetKwInOut{Output}{Output}
%	\Input{Hermitian matrix $\mb R \in \mathbb C^{K\times K}$; number of iterations $M\leq K$; starting values $\beta\pp{1}=0$, $\mb v\pp{0}=\mb 0_{K \times 1}$, unit-norm vector $\mb v\pp{1}$}
%	\Output{Eigenvalue estimates $[\lambda_1\pp{M}, \hdots,   \lambda_M\pp{M}]$}
%			\BlankLine
%		\For{iteration $j=1$ to $M$} {
%$	\mb q\pp{j} \la \mb R \mb v\pp{j}  - \beta\pp{j} \mb v\pp{j-1}$\; \label{Rv}
%$	\alpha\pp{j} \la {\mb q\pp{j}}^H \mb v\pp{j}$\; \label{alpha}
%$	\mb q\pp{j} \la \mb q\pp{j} - \alpha\pp{j} \mb v\pp{j}$\; \label{sub}
%$	\beta\pp{j+1} \la \| 	\mb q\pp{j}  \|$ \; \label{beta}
%$	\mb v\pp{j+1} \la \mb q\pp{j} / \beta\pp{j+1}$\; \label{vecv}
%	}
%	Construct $\mb T\pp{M}$ (\ref{T})\;
%	$[ \lambda_1\pp M, \hdots,  \lambda_M\pp M] \la $ eigenvalues of $\mb T\pp{M}$\;
%		\normalsize
%\end{algorithm}   

A more sophisticated eigenvalue algorithm is the LA, originally proposed by Lanczos in \cite{lan_orig}. 
The LA is applicable only to symmetric or Hermitian matrices (which is the case of the sample covariance matrix $\mb R$ considered here), but provides estimates of multiple eigenvalues (the number of estimated eigenvalues depends on the iterations of the algorithm) and has faster convergence than the PM \cite[Chapter 6]{book}.
The advantage of the LA over the PM lies in the fact that the LA takes into account, at every iteration $j$, the complete Krylov subspace
\beq
\mc K_j (\mb R, \mb v\pp 0) = \mathrm{span} \{ \mb v\pp 0, \mb R \mb v\pp 0, \hdots, \mb R^{j} \mb v\pp 0 \} 
\eeq
whereas the PM only considers the last term $\mb R^j \mb v\pp 0$.
The LA has been thoroughly studied by Paige \cite{paige72,paige76,paige80}. In particular, several computational variants
are compared  in \cite{paige72}, showing that some are numerically more stable than others.
Among the ``stabler'' variants, we adopt the one named ``A(1,7)'' in \cite{paige72} or equivalently ``A2'' in \cite{paige76}. 
The same version of the algorithm is presented in \cite[p. 651]{book2}. 
This variant is convenient in view of the decentralized implementation that will be developed in Section \ref{sec:dla}.

The derivation of the algorithm can be briefly outlined as follows. 
Due to the  Hermitian structure of  $\mb R$, for a given $M\leq K$, we can write
\beq
\mb R \mb V = \mb V \mb T
\eeq
where the columns of $\mb V \in \mathbb C^{K\times M}$  are mutually orthogonal unit-norm vectors and $\mb T \in \mathbb C^{M\times M}$ is a tridiagonal matrix.
If $M=K$, matrices  $\mb T$ and  $\mb R$ are similar, so their eigenvalues are the same. %        the eigenvalues of $\mb T$ are exactly those of $\mb R$, because the two matrices are similar.
However,  as Lanczos first noted, the eigenvalues of $\mb T$ (sometimes referred to as ``Ritz values'') turn out to be excellent approximations of the eigenvalues of $\mb R$ even when $M<K$. 
The LA is thus defined by iteratively equating the columns of $\mb R \mb V$ to those of $\mb V \mb T$. 
If we let
\beq
\label{T}
\mb T = \begin{bmatrix} 
\alpha\pp{1} & \beta\pp{2} & & & \\
\beta\pp{2} & \alpha\pp{2} & \beta\pp{3} & & \\
        & \cdots & \cdots & \cdots & \\
        &       & \beta\pp{M-1} & \alpha\pp{M-1} & \beta\pp{M}\\
        &  & & \beta\pp{M} & \alpha\pp{M}
 \end{bmatrix},
\eeq
the $j$-th iteration of the LA can be written as
\begin{align}
 \alpha\pp j & = \mb v\pp j^H \mb R \mb v\pp j \label{laa} \\
 \mb w\pp{j} &= \mb R \mb v\pp j -   \alpha\pp j \mb v \pp j - \beta\pp j \mb v\pp{j-1} \label{law}\\
 \beta\pp {j+1} &= \|\mb w\pp j \| \label{lab} \\
 \mb v\pp{j+1} &= \mb w\pp j /  \beta\pp {j+1}, \label{lav}
 \end{align}
 with an arbitrary starting vector $\mb v\pp 0$ of unit norm, and $\beta\pp 1 = 0$.
The above iteration is repeated for $j=1$ to $M$, thus obtaining the coefficients $\alpha\pp j$ and $\beta \pp j$ which are necessary to construct $\mb T$. 
The desired estimates $\hat \lambda_1, \hdots, \hat \lambda_M$ of the eigenvalues of $\mb R$ are then set to be the eigenvalues of  $\mb T$. Note that the eigenvalues of a tridiagonal matrix of size $M \times M$ can be efficiently computed with complexity $O(M^2)$
by using the spectral bisection method \cite{sb}.

%\section{Block Methods}
\section{Proposed Algorithms}
\label{algo}

We now investigate how the aforementioned PM and LA can be implemented in a distributed fashion over a wireless network. 
That is, the goal is for each node $k\in \{1, 2, \hdots, K\}$ to compute local estimates $\{\hat\lambda_i[k]\}$ of the eigenvalue(s) of interest, namely, $i=1$ for the PM and  $1\leq i \leq M$ for the LA.
Decentralized eigenvalue estimates should be as close as possible to their centralized counterparts: ideally, for every eigenvalue $\lambda_i$ of interest, we would like to have
\beq
\hat\lambda_i[k] = \hat\lambda_i \ \ \ \forall k.
\eeq

In the following sections we show that efficient DPM and DLA schemes can be developed by distributing the PM and LA vector iterations -- respectively, (\ref{pmiter}) and (\ref{laa})-(\ref{lav}) --  in such a way that the $k$-th element of vector $\mb v\pp j$, indicated by $v\pp j [k]$, is computed by node $k$. 
This is achieved by iterative exchange of messages between node $k$ and its neighbors, using AC routines. A similar principle was used in \cite{scaglione_ieee} in order to develop a distributed implementation of the Oja-Karhunen recursion for eigenvectors, as discussed also in Section \ref{intro}. 
Once the elements $v\pp j[k]$ are available at the $K$ nodes, inner product and norms necessary for eigenvalue computation are performed again by AC, while all element-wise operations (sums, multiplications by constants) are done locally at each node.
 Next, for both PM and LA, we first rewrite the global iteration in a way that is amenable to decentralized computation via AC, and then we break the global iteration down into a sequence of algorithmic steps to be executed by individual nodes.

\subsection{Decentralized Power Method}

The main result for the PM vector iteration is given by the following proposition.

\begin{proposition}
\label{prop1}
Given an ideal AC routine $\AC_\cdot(\cdot)$, the PM iteration (\ref{pmiter}) can be rewritten as 
\beq
\label{propdpm}
\mb v\pp{j+1} = \frac K N \diag \big\{  \mb Y \cdot  \big[ \AC_N \big(\Diag (\mb v\pp j ^* ) \cdot \mb Y \big) \big]^H  \big\}.
\eeq
\end{proposition}
\begin{IEEEproof}
From (\ref{pmiter}) we can write
\begin{align}
\mb v \pp{j+1} &= \frac 1 N \mb Y \mb Y^H \mb v\pp j \\
&= \frac 1 N \diag (\mb Y \mb Y^H \mb v\pp j \mb 1_K^T ) \\
&= \frac 1 N \diag \big(\mb Y \mb Y^H \Diag (\mb v\pp j) \mb 1 \mb 1^T\big) \\
&= \frac 1 N \diag \big[\mb Y     \big( \mb 1 \mb 1^T  \Diag (\mb v\pp j^*) \mb Y \big)^H    \big].  \label{proofdpm}
\end{align}
Now, if we let $\mb Z \triangleq \frac 1 K \mb 1 \mb 1^T  \Diag(\mb v\pp j^*) \mb Y$ and $\mb Z_0 \triangleq   \Diag(\mb v\pp j^*) \mb Y$, it is clear from (\ref{acdef}) that $\mb Z$ is the ideal AC output with $\mb Z_0$ as initial matrix:
\beq
\label{zz0}
\mb 1 \mb 1^T  \Diag(\mb v\pp j^*) \mb Y = K \cdot \AC_N \big( \Diag (\mb v\pp j^*) \mb Y \big).
\eeq 
Combining (\ref{proofdpm}) with (\ref{zz0}) yields (\ref{propdpm}).
\end{IEEEproof}

Complicated though it may look, expression (\ref{propdpm}) naturally leads to a decentralized implementation thanks to the following properties:
\textit{(i)} the $k$-th element of the output vector,  $v\pp{j+1}[k]$, is the $k$-th element of $\diag(\mb Y \mb Z^H)$, 
hence
\beq
v\pp{j+1}[k] = \mb z[k]^H \yr{k} 
\eeq
which can be computed locally by node $k$ (recall that $\yr{k} $ is the  vector of samples  received by node $k$ and $\mb z[k]$ is the $N$-dimensional AC output at node $k$);
%\beq
%v\pp{j+1}[k] = \sum_{n=1}^N 
%\eeq 
  \textit{(ii)} the input to $\AC_N(\cdot)$ is   such  that the  $k$-th  row  only contains node $k$'s local data $v\pp{j}[k]^* \cdot \yr{k}^T$ (recall that  $v\pp j[k]$ has been computed by node $k$ in the preceding iteration).

Assume now that the DPM iteration (\ref{propdpm}) has been repeated $M$ times.\footnote{Note that the number of algorithm iterations is denoted by $M$ because $M$ is in fact the dimension of the underlying Krylov subspace. This identity is more evident in the case of the LA.} The largest eigenvalue estimate $\hat \lambda_1$ (\ref{hatl1}) can be computed for all nodes by two additional calls to AC, as follows from the following proposition.

\begin{proposition}
\label{prop2}
Given an ideal AC routine $\AC_\cdot(\cdot)$, after $M$ iterations of (\ref{propdpm}), $K$ local copies of the largest eigenvalue estimate (\ref{hatl1}) can be obtained as
\beq
\label{propl1}
\hat{\lambda}_1 \mb 1_K =  \frac K N   \rnorm   \big[ \AC_N \big(\Diag(\mb v\pp M ^*) \cdot \mb Y \big)  \big]    \ewfrac  \AC_1 (\mb v\pp M^* \ewtimes \mb v\pp{M} )  .
\eeq 
where $\mathtt{RN}: \mathbb C^{K\times m} \to  \mathbb C^K$ is a function that returns the squared $\ell^2$ norms of  the rows of an input matrix (see Tab. \ref{notation}).
\end{proposition}
\begin{IEEEproof}
We first write the global output for all $K$ nodes after $M$ iterations as
\beq
\label{p2pre}
\hat \lambda_1 \mb 1_K =     \big( \mb 1_K \mb v\pp M^H \mb R \mb v\pp{M}  \big)  \ewfrac \big( \mb 1_K \mb v\pp M^H \mb v\pp{M}  \big),
\eeq 
where we have used (\ref{hatl1}) and applied element-wise division.
The numerator can be written as
\begin{align}
\mb 1 \mb v\pp M^H \mb R \mb v\pp{M}   &= \frac 1 N   \mb 1 \mb v\pp M^H \mb Y \mb Y^H \mb v\pp{M} \label{p2numi}  \\
&= \frac 1 N \mb 1  \big\| \mb v\pp M^H \mb Y\big\|^2 \\
&= \frac 1 N \mb 1 \big\| \mb 1^T \Diag ( \mb v\pp M^*) \mb Y \big\|^2  \\
&=\frac 1 N  \rnorm \big[ \mb 1 \mb 1^T \Diag ( \mb v\pp M^*) \mb Y \big] \\
&= \frac {K^2} N \rnorm \big[   \AC_N\big(\Diag (\mb v\pp M ^*)  \mb Y \big)\big].  \label{p2num}
\end{align}
For the denominator, we can write
\begin{align}
\mb 1  \mb v\pp M^H \mb v\pp{M}
&=  \mb 1 \mb 1^T  ( \mb v\pp M^* \ewtimes \mb v\pp{M} )  \label{p2deni} \\
&= K \cdot \AC_1 (\mb v\pp M^* \ewtimes \mb v\pp{M} ).   \label{p2den}
\end{align}
Combining (\ref{p2num}) with (\ref{p2den}) and simplifying $K$ finally yields (\ref{propl1}).
\end{IEEEproof}

%
%\begin{proposition}
%\label{prop2}
%Given an ideal AC routine $\AC_\cdot(\cdot)$, after $M$ iterations of (\ref{propdpm}), $K$ copies of the largest eigenvalue estimate (\ref{hatl1}) can be obtained as
%\beq
%\label{propl1}
%\hat{\lambda}_1 \mb 1_K = \left[ \AC_1\left(\mb v\pp M^* \ewtimes \mb v\pp{M+1} \right) \right] \ewfrac \left[ \AC_1\left(\mb v\pp M^* \ewtimes \mb v\pp{M} \right) \right].
%\eeq 
%\end{proposition}
%\begin{IEEEproof}
%We first note  that $\mb R \mb v\pp M = \mb v\pp {M+1}$, which has been computed in the $M$-th iteration of (\ref{propdpm}). 
%Hence, (\ref{hatl1}) after iteration $M$ can be written as
%\begin{align}
%\hat\lambda_1 &= \frac{\mb v\pp M^H \mb v\pp{M+1}}{\mb v\pp M^H \mb v\pp{M}} \\
%&= \frac {\mb 1^T \left( \mb v\pp M^* \ewtimes \mb v\pp{M+1} \right) } { \mb 1^T  \left(  \mb v\pp M^* \ewtimes \mb v\pp{M} \right)}. \label{qqq}
%\end{align}
%Now, if we multiply both sides of (\ref{qqq}) by $\mb 1_K$, we obtain
%\begin{align}
%\mb 1 \hat \lambda_1 &=  \left[\mb 1 \mb 1^T \left( \mb v\pp M^* \ewtimes \mb v\pp{M+1} \right) \right] \ewfrac \left[ \mb 1 \mb 1^T  \left(  \mb v\pp M^* \ewtimes \mb v\pp{M} \right) \right],
%\end{align}
%which yields (\ref{propl1}) by applying the definition of AC (\ref{acdef}).
%\end{IEEEproof}

Similar to (\ref{propdpm}), expression (\ref{propl1}) can be readily implemented in a distributed manner.
At the numerator, every node $k$ computes an AC vector $\mb z[k]\in \mathbb C^N$ starting from the initial value $v\pp{M}[k]^* \cdot \yr{k}^T$, just like in the vector iteration, and takes the  norm of $\mb z[k]$ locally.
At the denominator, the scalar AC input at node $k$ is simply the local quantity $|v\pp M[k]|^2$. 
Element-wise division is then performed internally by each node.

Thanks to the results of Propositions \ref{prop1} and \ref{prop2}, and replacing the ideal AC routine by one with finite averaging time $t$, the DPM can be written in algorithmic form as summarized in Alg. \ref{dpm}. The averaging time or number of iterations $t$ is assumed to be either predefined or adjusted online at every iteration, based on the starting vector and a target error bound. 
%Note that the sequence of steps in Alg. \ref{dpm} is to be executed by all nodes $k\in \{1, \hdots, K\}$ in parallel.
%For sake of generality, the  running time of the AC routine is indicated as $t_j$, because it may be adjusted for different DPM iterations $j$ (see also Section \ref{error_anal}). 

\begin{algorithm}[h!]
\ifsmallalg \small \fi
	\caption{Decentralized power method}
	\label{dpm}
	\SetKwInOut{Input}{Input}\SetKwInOut{Output}{Output}
	\Input{Received signal vectors $\yr{k}$ $\in \mathbb C^{N}$ for  $k\in \{1, \hdots, K\}$; number of iterations $M$; starting values $v\pp{0}[k]$ $\forall k$; averaging time $t$.} %sequence of AC times $\{t_1, \hdots, t_M\}$}
	\Output{Eigenvalue estimates   $\hat\lambda_{1}[k]$ $\forall k$.}
			\BlankLine
					\For {all nodes $k$ in parallel}{
		\For{iteration $j=1$ to $M$} {
		$\mb z[k]^T =  \AC^{t}_N[k] \left(v\pp{j-1}[k]^* \cdot \yr{k}^T \right)$\;  \label{dpm1}  % \zr{0}{k} =
		Compute locally $v\pp{j}[k] = \frac K N  \mb z[k]^H \yr{k}$\; \label{dpmloc1}
		}
		$\mb z[k]^T = \AC^{t}_N[k] \left(v\pp{M}[k]^* \cdot \yr{k}^T \right)$\; \label{dpm2}
		$d[k] = \AC^t_1[k]\left( |v\pp M[k]|^2 \right)$\;  \label{dpm3}
		Compute locally $\hat\lambda_1[k] = \frac K N \cdot {\| \mb z[k]\|^2 }/{ d[k]}$\;   \label{dpml}
		}
		\normalsize
\end{algorithm}  

\subsection{Decentralized Lanczos Algorithm}
\label{sec:dla}

Consider now the LA iteration (\ref{laa})-(\ref{lav}). We note that (\ref{laa}) has the same structure as the numerator of (\ref{hatl1}), and (\ref{law}) is similar to the PM iteration (\ref{pmiter}), with additional terms  which can  be computed locally provided that each node $k$ has stored a local copy of $\alpha\pp j$, $\beta\pp j$, and of the $k$-th element of vectors $\mb v\pp j$, $\mb v\pp{j-1}$.
Then, the normalization step (\ref{lab})-(\ref{lav})  is similar to the denominator  of (\ref{hatl1}), therefore it can be implemented by scalar AC and element-wise division.
Based on the above observations, we can state the following result.

\begin{proposition}
\label{prop3}
Given an ideal AC routine $\AC_\cdot(\cdot)$, the LA (\ref{laa})-(\ref{lav}) can be rewritten as
\ifdouble
\begin{align}
\alpha\pp j \mb 1_K &= \frac {K^2} N \rnorm \big[  \AC_N\left(\Diag (\mb v\pp j ^*) \cdot \mb Y \right) \big]  \label{dlaa} \\ 
 \mb w\pp{j} &= \frac K N \diag \big\{  \mb Y \cdot \big[ \AC_N \big(\Diag (\mb v\pp j ^*) \cdot \mb Y \big) \big]^H  \big\} \nonumber \\
   & \ \ \ -\alpha\pp j \mb v\pp j - \beta\pp j \mb v\pp{j-1} \label{dlaw}  \\
 \beta^2 \pp {j+1} \mb 1_K &= K \cdot \AC_1 \big(\mb w\pp j^* \ewtimes \mb w\pp j \big) \label{dlab}  \\
 \mb v\pp{j+1} &= \mb w\pp j / \beta\pp{j+1} \label{dlav} .
\end{align}
\else
\begin{align}
\alpha\pp j \mb 1_K &= \frac {K^2} N \rnorm \big[  \AC_N\left(\Diag (\mb v\pp j ^*) \cdot \mb Y \right) \big]  \label{dlaa} \\ 
 \mb w\pp{j} &= \frac K N \diag \big\{  \mb Y \cdot \big[ \AC_N \big(\Diag (\mb v\pp j ^*) \cdot \mb Y \big) \big]^H  \big\} -\alpha\pp j \mb v\pp j - \beta\pp j \mb v\pp{j-1} \label{dlaw}  \\
 \beta^2 \pp {j+1} \mb 1_K &= K \cdot \AC_1 \big(\mb w\pp j^* \ewtimes \mb w\pp j \big) \label{dlab}  \\
 \mb v\pp{j+1} &= \mb w\pp j / \beta\pp{j+1} \label{dlav} .
\end{align}
\fi
\end{proposition}
\begin{IEEEproof}
The proof is a combination of the same steps  already used in Prop. 1 and Prop. 2.
More precisely, 
(\ref{dlaa}) follows from (\ref{laa}) using (\ref{p2numi})-(\ref{p2num}); 
 (\ref{dlaw}) from (\ref{law}) using (\ref{propdpm});
 (\ref{dlab}) from (\ref{lab}) using (\ref{p2deni})-(\ref{p2den});
and  (\ref{dlav}) is simply  (\ref{lav}).
\end{IEEEproof}
The above result can be mapped to the  decentralized algorithm reported in Alg. \ref{dla}, where a  realistic AC scheme $\AC^t_\cdot$ is adopted.
%We note that, according to \cite{paige76}, the square root and the normalization steps (\ref{dlab})-(\ref{dlav}) ``are not essential, but they prevent overflow and underflow''. We have included these steps, in spite of the resulting complexity increase, because otherwise the algorithm would become too sensitive to numerical problems. 

  \begin{algorithm}[h!]
\ifsmallalg \small \fi
	\caption{Decentralized Lanczos algorithm}
	\label{dla}
	\SetKwInOut{Input}{Input}\SetKwInOut{Output}{Output}
	\Input{	Received signal vectors $\yr{k}$ $\in \mathbb C^{N}$ for  $k\in \{1, \hdots, K\}$; number of iterations $M\leq K$; starting values $v\pp{1}[k]$ $\forall k$, such that $\sum_{k=1}^K |v\pp{1}[k]|^2 = 1$, $\beta\pp 1[k] = 0$ $\forall k$; averaging time $t$.}
	\Output{Eigenvalue estimates  $\{ \hat \lambda_{1}[k] , \hdots,   \hat \lambda_M[k] \}$ $\forall k$.}
			\BlankLine
		\For{all nodes $k$ in parallel}{
				\For{iteration $j=1$ to $M$} {
					$\mb z[k]^T =  \AC^{t}_N[k] \left(v\pp{j}[k]^* \cdot \yr{k}^T \right)$\;  \label{dla1} % \zr{0}{k} =
		Compute locally $\alpha\pp{j}[k] = \frac {K^2}{N}  \|\mb z[k]\|^2$  \;  \label{dlanorm}
				Compute locally $w\pp j[k] = \frac K N  \mb z[k]^H \yr{k} - \alpha\pp j[k] \cdot v\pp j[k] - \beta\pp j[k] \cdot v\pp{j-1}[k]$\;  \label{dlawk}
				$b[k] = \AC^t_1[k] (|w\pp j[k]|^2)$\;  \label{dla2}
				Compute locally $\beta\pp{j+1}[k] = \sqrt{K\cdot b[k]}$ and $v\pp{j+1}[k] = w\pp j[k]/\beta\pp{j+1}[k]$;  \label{dlabk}
%			  $\beta\pp{j+1}[k] = \sqrt{K\cdot \AC^t_1[k] (|w\pp j[k]|^2)}$ and $v\pp{j+1}[k] = w\pp j[k]/\beta\pp{j+1}[k]$;
						}
		Construct locally $\mb T[k]$ from (\ref{T}) using $\alpha\pp{1}[k], \hdots, \alpha\pp{M}[k]$, $\beta\pp{2}[k], \hdots, \beta\pp{M}[k]$\; \label{dlaTk1}
		Compute locally $\{\hat\lambda_1[k], \hdots, \hat\lambda_M[k]\} = $ eigenvalues of  $\mb T[k]$\;  \label{dlaTk}
			}
		\normalsize
\end{algorithm}

\section{Error and Complexity Analysis}
\label{perf}

In this section we analyze the impact of non-ideal AC algorithms on the error and numerical complexity (especially in terms of network signaling) for the DPM and the DLA.

\subsection{DPM Error}

The DPM algorithm (Alg. \ref{dpm}) involves three sources of error due to AC.
We define the first error term as
\ifdouble
\begin{align}
\mb E\pma \pp j & \triangleq 
\AC^t_N  \big(\Diag (\mb v\pp {j-1} ^*) \cdot \mb Y \big) - \frac 1 K \mb 1 \mb 1^T \Diag (\mb v\pp {j-1} ^* ) \cdot \mb Y  \nonumber \\
 &\in \mathbb C^{K\times N},  \label{epm1}
\end{align}
\else
\beq
\label{epm1}
\mb E\pma \pp j \triangleq 
\AC^t_N  \big(\Diag (\mb v\pp {j-1} ^*) \cdot \mb Y \big) - \frac 1 K \mb 1 \mb 1^T \Diag (\mb v\pp {j-1} ^* ) \cdot \mb Y  
 \in \mathbb C^{K\times N},
\eeq
\fi
which occurs at every iteration of (\ref{propdpm}), corresponding to line \ref{dpm1} of Alg. \ref{dpm}.  Following the previously used convention, we denote by $\mb e\pma \pp j [k]^T$ the $k$-th row of $\mb E\pma \pp j$.
The second and third error terms arise from (\ref{propl1}), i.e., lines \ref{dpm2} and \ref{dpm3} of the algorithm, 
and are defined as
\ifdouble
\begin{align}
 \mb E\pmb &\triangleq 
 \AC^t_N \big( \Diag (\mb v\pp M ^*) \cdot \mb Y \big) - \frac 1 K \mb 1 \mb 1^T \Diag (\mb v\pp M ^*) \cdot \mb Y \nonumber \\
  &\in \mathbb C^{K\times N} , \label{epm2} \\
  \mb e\pmc  &\triangleq 
  \AC^t_1 (\mb v\pp M^* \ewtimes \mb v\pp{M} )  - \frac 1 K \mb 1 \mb 1^T (\mb v\pp M^* \ewtimes \mb v\pp{M} ) \nonumber \\
   &\in \mathbb C^{K}.   \label{epm3}
\end{align}
\else
\begin{align}
 \mb E\pmb &\triangleq 
 \AC^t_N \big( \Diag (\mb v\pp M ^*) \cdot \mb Y \big) - \frac 1 K \mb 1 \mb 1^T \Diag (\mb v\pp M ^*) \cdot \mb Y 
  \in \mathbb C^{K\times N}, \label{epm2} \\
  \mb e\pmc  &\triangleq 
  \AC^t_1 (\mb v\pp M^* \ewtimes \mb v\pp{M} )  - \frac 1 K \mb 1 \mb 1^T (\mb v\pp M^* \ewtimes \mb v\pp{M} ) 
   \in \mathbb C^{K}.   \label{epm3}
\end{align}
\fi
Again, we refer to the $k$-th row of  $\mb E\pmb $ as $\mb e\pmb \pp j [k]^T$ 
and to the $k$-th element of $\mb e\pmc $ as $e\pmc [k]$.
Note that the three above defined errors   are all instances of the general formula (\ref{acdef}) applied with different inputs. To simplify the notation, we have dropped subscript $t$ in the symbols of error variables.

With regard to the DPM convergence, the most important term is clearly $\mb E\pma$, because this error is added at each iteration of the algorithm. 
The impact of  $\mb E\pma$ on the evolution of PM vectors $\mb v\pp j$ is expressed by the following result.

\begin{proposition}
\label{prop4}
Given a non-ideal AC scheme that introduces an error term $\mb E\pma\pp j$ as defined in (\ref{epm1}), the resulting DPM vector after $M$ iterations can be written as
\beq
\label{prop4th}
\mb v\pp M = \mb R^M \mb v\pp 0 + \frac 1 N \sum_{j=1}^{M} \mb R^{M-j} \diag \big[  \mb Y \big(\mb E\pma\pp j\big)^H \big] .
\eeq
\end{proposition}
\begin{IEEEproof}
By applying the same steps as in the proof of Prop. \ref{prop1} with the AC function defined in (\ref{acdef}) with $\mb E_t$ replaced by $\mb E\pma\pp {j}$, we can write for  any iteration $j$
\beq
\label{prop4j}
\mb v\pp {j} = \mb R \mb v\pp {j-1} +  \frac 1 N  \diag \big[  \mb Y \big(\mb E\pma\pp j\big)^H \big] .
\eeq  
The above formula, applied recursively for $M$ iterations,  yields (\ref{prop4th}).
\end{IEEEproof}
The term $ \mb R^M \mb v\pp 0$ in (\ref{prop4th}) represents the ideal PM output, while the summation on the r.h.s. represents the error. 
For brevity, we define  
\beq
\mb d\pp j \triangleq  \frac 1 N  \diag \big[  \mb Y \big(\mb E\pma\pp j\big)^H \big] ,
\eeq
which represents the error introduced by AC in a single iteration $j$. Its $k$-th element (i.e., the component for node $k$) is $d\pp j[k] = \frac 1 N  (\mb e\pp j[k])^H \yr{k}$.

% New part on convergence

%\footnote{With reference to the signal model of Section \ref{app} (typically used in the spectrum sensing literature), the property $\lambda_1>1$ can be simply enforced by normalizing the received signal samples $\mb Y$ by the noise standard deviation or an upper bound thereof, so that the normalized noise variance is  $\sigma^2 \geq 1$.  As such, assuming noise-only hypothesis, the eigenvalues of the sample covariance matrix for finite $K,N$ are distributed around the asymptotic value $\sigma^2$ (to which all eigenvalues converge as $N \to \infty$) \cite{anderson}, hence $\lambda_1 > \sigma ^2 \geq 1$.  
% Furthermore, in the asymptotical regime $K,N \to \infty$ with finite $K/N \triangleq \gamma$, the largest eigenvalue converges almost surely to the upper limit of the Mar{\v c}enko-Pastur bound $\sigma^2 (1+\sqrt \gamma) > 1$ \cite{rmt}. 
%The property $\lambda_1>1$ is satisfied once more if signals are present in addition to noise, resulting in eigenvalues strictly larger than those of the noise-only case \cite{baik}.
%}

Now, the convergence of the DPM depends on the relative magnitude of the error term $\sum_{j=1}^{M} \mb R^{M-j} \mb d\pp j$ compared to the desired term $\mb R^M \mb v\pp 0$ as $M\to \infty$ (recall that both terms are unnormalized).
Let us assume that the spectral radius of $\mb R$ (i.e., $\lambda_1$, since the eigenvalues are real and positive) is larger than $1$.\footnote{This assumption simplifies the mathematical analysis, and it is not a limitation in practice, because it is always possible to rescale the data samples $\mb Y$ such that $\lambda_1>1$.}
Then, the DPM error converges asymptotically to zero if the magnitude of the AC error vector grows slower than a certain rate.
%Under this assumption, it can be shown that the error vectors $\mb d \pp j$ do not affect the asymptotic convergence of the vector $\mb v\pp M$ to the true eigenvector $\mb u_1$, as long as the norm of such errors grows  at a rate slower than $\lambda_2^M$.  
%More precisely, the rate of convergence is still $O((\lambda_2/\lambda_1)^M)$, and the perturbations introduced by the errors $\mb d \pp j$ are asymptotically negligible.
The result is expressed formally by the following proposition.

\begin{proposition}
\label{convdpm}
Let $\theta\pp j \in [0, \pi/2]$ be the angle between the true eigenvector and its DPM estimate $\mb v\pp j$ after $j$ iterations,
defined by
\beq
\cos \theta\pp j = \mb u_1 ^H \frac{\mb v\pp j}{ \| \mb v\pp j \| },   \ \  0 \leq j \leq M,
\eeq 
and assume  $\cos \theta\pp 0 \neq 0$, $\lambda_1>1$. \ifdouble \else \\ \fi
 Then, asymptotically in $M$, provided that $\|\mb d\pp M\|_\infty = o\left( \frac{1}{M+1} \left( \frac{\lambda_1} {\max\{\lambda_2, 1  \}}  \right)^M  \right)$, we have
\beq
\lim_{M\to \infty}
|\sin \theta\pp M| = 0.
\eeq
\end{proposition}

\begin{IEEEproof}
Similarly as in \cite[p. 406]{golub}, we start by expressing vectors $\mb v\pp 0$ and $\mb d\pp j$ in the eigenbasis $(\mb u_1, \hdots, \mb u_K)$, so that
\begin{align}
\mb v \pp 0 &= a_{0,1} \mb u_1 + \hdots + a_{0,K} \mb u_K, \\
\mb d\pp j &= a_{j,1} \mb u_1 + \hdots + a_{j,K} \mb u_K, \ \ 1\leq j \leq M. \label{aji}
\end{align}
%for all $1\leq j \leq M$. 
By assumption we have $|a_{0,1}| =\cos \theta\pp 0 \neq 0$.
The DPM vector after $M$ iterations can be then written as
\begin{align}
\mb v\pp M &= \mb R^M \mb v\pp 0 + \sum_{j=1}^{M} \mb R^{M-j} \mb d\pp j \\
&=  \sum_{j=0}^M a_{j,1} \lambda_1 ^{M-j}   \mb u_1 + \hdots +
   \sum_{j=0}^M a_{j,K} \lambda_K ^{M-j}  \mb u_K,
\end{align}
and consequently
\begin{align}
|\sin\theta\pp M|^2 &= 1 - \frac{ \left|\mb u_1^H \mb v\pp M   \right|^2      }{\| \mb v\pp M \|^2}  \\
&=  1 - \frac{    \left|  \sum_{j=0}^M a_{j,1} \lambda_1 ^{M-j}    \right|^2      } {  \sum_{i=1}^K   \left|  \sum_{j=0}^M a_{j,i} \lambda_i ^{M-j} \right|^2  }  \\
&=  \frac{      \sum_{i=2}^K   \left|  \sum_{j=0}^M a_{j,i} \lambda_i ^{M-j} \right|^2       }{       \sum_{i=1}^K   \left|  \sum_{j=0}^M a_{j,i} \lambda_i ^{M-j} \right|^2     } \\
& \leq \frac        {      \sum_{i=2}^K   \left|   \sum_{j=0}^M a_{j,i} \lambda_i ^{M-j} \right|^2       }        {    \left|  \sum_{j=0}^M a_{j,1} \lambda_1 ^{M-j}    \right|^2    } .  \label{qq1}
\end{align}
By letting $a_i(M) \triangleq \max_{0 \leq j \leq M} |a_{j,i}|$ and dividing numerator and denominator by $\lambda_1^{2M}$, we have
\begin{align}
\mbox{(\ref{qq1})} &\leq   \frac        {      \sum_{i=2}^K   \left| a_i(M)  \sum_{j=0}^M  \lambda_i ^{M-j} \right|^2       }        {    \left|  \sum_{j=0}^M a_{j,1} \lambda_1 ^{M-j}    \right|^2    } \\
 &= \frac{  \sum_{i=2}^K  a_i^{2M}   \left| \sum_{j=0}^M ( {\lambda_i}/{\lambda_1} )^M \lambda_i^{-j}  \right|^2   }     {  \left|  \sum_{j=0}^M a_{j,1} \lambda_1 ^{-j}    \right|^2   }.
 \label{qq2}
\end{align}
Let us first consider the denominator, which can be written as 
$\left|  \sum_{j=0}^M a_{j,1} \overline \lambda_1 ^{j}    \right|^2$, with $ \overline \lambda_1 \triangleq 1/\lambda_1 \in (0,1)$. 
We have to consider two cases. 
\textit{(i)}  Assume the series is absolutely convergent, so that $\lim_{M \to \infty} \sum_{j=0}^M |a_{j,1}|<\infty$. Moreover, since $a_{0,1}\neq 0$, we have $\lim_{M \to \infty} \sum_{j=0}^M |a_{j,1}| = \alpha $ for some $\alpha>0$. Now note that  $ \sum_{j=0}^M a_{j,1} \overline \lambda_1 ^{j}   $     is the Z-transform of the sequence $\{a_{j,1}\}_{j=0}^{\infty}$ at $\overline \lambda_1$. So, as $\overline \lambda_1 < 1$, the Z-transform exists, and therefore the series converges to a value $\beta$.
\textit{(ii)} Assume now that $\lim_{M \to \infty} \sum_{j=0}^M |a_{j,1}|= \infty$. In this case, the radius of convergence of the power series in $\overline \lambda_1$ is $0$, which means that the series diverges for any value of $\overline \lambda_1$. 
Combining the two cases, we conclude that $\lim_{M \to \infty} \left|  \sum_{j=0}^M a_{j,1}  \lambda_1 ^{-j}    \right|^2 \in [\beta^2, \infty]$, which means that in the worst case the denominator converges to a finite positive value.

Consider now the numerator, which we can write as 
\beq 
A_M \triangleq  \sum_{i=2}^K  a_i(M)^{2}  \underbrace{ \left| \sum_{j=0}^M ( {\lambda_i}/{\lambda_1} )^M \lambda_i^{-j}  \right|^2}_{\triangleq A_M\ap i}.  
\eeq
Again, we have two cases. 
\textit{(i)} If $\lambda_i \geq 1$, then $\lambda_i ^j \geq 1 $ for any $0 \leq j \leq M$, hence 
\beq
A_M \ap i \leq \left| \sum_{j=0}^M (\lambda_i/ \lambda_1)^M \right|^2 = \left| (M+1)  (\lambda_i/ \lambda_1)^M \right|^2,
\eeq
and, recalling that $\lambda_i < \lambda_1$ for any $i\geq 2$ and  applying L'Hopital's rule, we obtain $\lim_{M \to \infty} A_M \ap i = 0$.
\textit{(ii)} If $\lambda_i < 1$, we have
\beq
A_M \ap i = \left| \sum_{j=0}^M  \lambda_1^{-M} \lambda_i^{M-j} \right|^2 \leq \left| \sum_{j=0}^M   \lambda_1^{-M} \right|^2 = \left| (M+1)  \lambda_1^{-M}  \right|^2, 
\eeq
whose limit is again $0$.
Combining these results yields
\beq
A_M \leq \sum_{i=2}^K a_i(M) ^2\left|  (M+1) \left( \frac{\max\{\lambda_2, 1  \}}{\lambda_1}  \right)^M \right|^2.
\eeq
Now suppose that $\max_{2\leq i\leq K} a_i(M) \leq \gamma(M)$. Then, the numerator converges to $0$ if $\gamma(M) = o\left( \frac{1}{M+1} \left( \frac{\lambda_1} {\max\{\lambda_2, 1  \}}  \right)^M  \right) $.
The fact that $\max_{2\leq i\leq K} a_i(M)$ is equal (up to a constant) to $\|\mb d\pp M\|_\infty$ %(and, by equivalence of norms,  to any other norm of $\mb d\pp M$ up to a constant) 
completes the proof.
\end{IEEEproof}

We next consider the impact of error terms $\mb E\pmb$ and  $\mb e\pmc$, which  only concern the eigenvalue computation phase at the final iteration. Let $\enu[k]$ and $\ede[k]$ be the errors introduced by non-ideal AC in step \ref{dpml} of Alg. \ref{dpm}, defined such that  
\beq
\label{l1err}
\hat\lambda_1[k] = %\frac K N \cdot 
 \frac{ \mb v\pp{M}^H \mb R \mb v\pp{M} + \enu[k]} {\mb v\pp{M}^H \mb v\pp{M} + \ede[k] }.
\eeq
The following proposition provides the values of $\ede[k]$ and $\enu[k]$ as a function of  $\mb E\pmb$ and  $\mb e\pmc$.

\begin{proposition}
\label{prop5}
Given a non-ideal AC scheme that introduces error terms $\mb E\pmb$ and $\mb e\pmc$  defined in  (\ref{epm2}) and (\ref{epm3}), the resulting errors in $\hat \lambda_1[k]$ are given by
\ifdouble
\begin{align}
\label{prop5th1}
\enu[k] &= \frac K N \big[ (\mb e\pmb[k])^T \mb Y^H \mb v\pp M + \mb v\pp M^H \mb Y (\mb e\pmb[k])^* \big] \nonumber \\
 & \ \ \ + \frac {K^2} N \big\|  \mb e\pmb[k] \big\|^2, \\
\ede[k] &= K \cdot e\pmc[k].
\label{prop5th2}
\end{align}
\else
\begin{align}
\label{prop5th1}
\enu[k] &= \frac K N \big[ (\mb e\pmb[k])^T \mb Y^H \mb v\pp M + \mb v\pp M^H \mb Y (\mb e\pmb[k])^* \big] + \frac {K^2} N \big\|  \mb e\pmb[k] \big\|^2, \\
\ede[k] &= K \cdot e\pmc[k].
\label{prop5th2}
\end{align}
\fi
%where $ (\mb e\pmb[k])^T$ and $e\pmc[k]$ are, respectively, the $k$-th row of $\mb E\pmb$ and the $k$-th element of vector $\mb e\pmc$.
\end{proposition}
\begin{IEEEproof}
Using the results of Prop. \ref{prop2} and introducing non-ideal AC, we can write
\ifdouble
\begin{align}
\hat\lambda_1[k] &= \frac K N \cdot \frac{\big\| \frac 1 K \mb v\pp M^H \mb Y +  (\mb e\pmb[k])^T \big\|^2 }{\frac 1 K  \|\mb v \pp M\|^2 + e\pmc[k]} \\
&= \frac{1}{ \|\mb v \pp M\|^2 + K   e\pmc[k] }  \Big\{      \tfrac 1 N  \big\|\mb v\pp M^H \mb Y   \big\|^2  +     \tfrac {K^2} N \|  \mb e\pmb[k] \|^2
  \nonumber \\
   & \ \ \   +    \tfrac K N \big[ (\mb e\pmb[k])^T \mb Y^H \mb v\pp M + \mb v\pp M^H \mb Y (\mb e\pmb[k])^* \big]    \Big\}.
     \label{proof5end}
\end{align}
\else
\begin{align}
\hat\lambda_1[k] &= \frac K N \cdot \frac{\big\| \frac 1 K \mb v\pp M^H \mb Y +  (\mb e\pmb[k])^T \big\|^2 }{\frac 1 K  \|\mb v \pp M\|^2 + e\pmc[k]} \\
&= \frac{ \frac 1 N  \big\|\mb v\pp M^H \mb Y   \big\|^2  +   \frac K N \big[ (\mb e\pmb[k])^T \mb Y^H \mb v\pp M + \mb v\pp M^H \mb Y (\mb e\pmb[k])^* \big] + \frac {K^2} N \|  \mb e\pmb[k] \|^2 } 
{ \|\mb v \pp M\|^2 + K   e\pmc[k] }.  \label{proof5end}
\end{align}
\fi
Since $ \frac 1 N  \big\|\mb v\pp M^H \mb Y   \big\|^2 = \mb v\pp{M}^H \mb R \mb v\pp{M}$, we note that (\ref{proof5end}) is equivalent to (\ref{l1err}), with error expressions  $\enu[k]$ and $\ede[k]$ given, respectively, by (\ref{prop5th1}) and (\ref{prop5th2}).
\end{IEEEproof}

By simple algebraic manipulations, and letting
$\hat \lambda_1\ide \triangleq  \big( \mb v\pp{M}^H \mb R \mb v\pp{M} \big) / \| \mb v\pp{M} \|^2    $ be the  estimate of $\lambda_1$ at the $M$-th DPM iteration without eigenvalue computation errors, we can write
\beq
\hat\lambda_1[k] = \frac{1}{1 +  {\ede[k]}/{\|\mb v\pp M\|^2}}   \cdot
 \hat \lambda_1\ide + 
 \frac{\enu[k]}{\|\mb v\pp M\|^2 + \ede[k]} .
\eeq
The above expression shows immediately that the eigenvalue computation error becomes asymptotically negligible, as $ \| \mb v\pp{M} \|^2 \to \infty$ as $M \to \infty$.

%By applying the triangle inequality, the error at the numerator can be bounded by
%\beq
%|\enu[k]|
%\eeq

%\[
%\diag(\mb Y \mb Z^H) = 
%%\]
%\beq
%\label{hatl1}
%\hat \lambda_1 =   \dfrac{ \mb v\pp{M}^H \mb R \mb v\pp{M}} {\mb v\pp M^H \mb v\pp{M}}.
%\eeq
%
%\beq
%	\hat\lambda_1[k] =	\frac{ \left\| \AC^{t}_N[k] \left(v\pp{M}[k]^* \cdot \yr{k}^T \right)\right\|^2 / N  }
%		{ \AC^t_1[k]\left( |v\pp M[k]|^2 \right)}
%\eeq

\subsection{DLA Error}
\label{perf_dla}

%
%The DPM algorithm (Alg. \ref{dpm}) involves three sources of error due to AC.
%We define the first error term as
%\beq
%\label{epm1}
%\mb E\pma \pp j \triangleq 
%\AC^t_N\left(\Diag\left(\mb v\pp {j-1} ^*\right) \cdot \mb Y\right) - \frac 1 K \mb 1 \mb 1^T \Diag\left(\mb v\pp {j-1} ^*\right) \cdot \mb Y  
% \in \mathbb C^{K\times N},
%\eeq
%which occurs at every iteration of (\ref{propdpm}), corresponding to line \ref{dpm1} of Alg. \ref{dpm}.  Following the usual convention, we denote by $\mb \mb e\pma \pp j [k]^T$ the $k$-th row of $\mb E\pma \pp j$.
%The second and third error terms arise from (\ref{propl1}), i.e., lines \ref{dpm2} and \ref{dpm3} of the algorithm, 
%and are defined as
%\begin{align}
% \mb E\pmb &\triangleq 
% \AC^t_N\left(\Diag\left(\mb v\pp M ^*\right) \cdot \mb Y\right) - \frac 1 K \mb 1 \mb 1^T \Diag\left(\mb v\pp M ^*\right) \cdot \mb Y
%  \in \mathbb C^{K\times N} \label{epm2} \\
%  \mb e\pmc  &\triangleq 
%  \AC^t_1\left(\mb v\pp M^* \ewtimes \mb v\pp{M} \right)  - \frac 1 K \mb 1 \mb 1^T \left(\mb v\pp M^* \ewtimes \mb v\pp{M} \right) 
%   \in \mathbb C^{K}.   \label{epm3}
%\end{align}

The DLA  involves two error terms due to AC, in lines \ref{dla1} and \ref{dla2} of Alg. \ref{dla}.
The first error  arises from
in (\ref{dlaa})-(\ref{dlaw}) and is defined as 
\ifdouble
\begin{align}
\mb E\laa \pp j &\triangleq 
\AC^t_N \big(\Diag (\mb v\pp {j} ^*) \cdot \mb Y \big) - \frac 1 K \mb 1 \mb 1^T \Diag (\mb v\pp {j} ^*) \cdot \mb Y   \nonumber \\
 &\in \mathbb C^{K\times N}.  \label{ela1}
\end{align}
\else
\beq
\label{ela1}
\mb E\laa \pp j \triangleq 
\AC^t_N \big(\Diag (\mb v\pp {j} ^*) \cdot \mb Y \big) - \frac 1 K \mb 1 \mb 1^T \Diag (\mb v\pp {j} ^*) \cdot \mb Y  
 \in \mathbb C^{K\times N}.
\eeq
\fi
%in analogy with $\mb E\pma\pp j$ in the DPM (\ref{epm1}).
%The first error, denoted by 
%\beq
%\label{ela1}
%\mb E\laa \pp j 
%= \big[ \mb e \laa \pp j[1], \hdots, \mb e \laa \pp j[K] \big]^T 
%\in \mathbb C^{K\times N},
%\eeq 
The second error originates from (\ref{dlab}) and is defined as
\beq
\label{ela2}
\mb e\lab \pp j \triangleq 
  \AC^t_1\big(\mb w\pp j^* \ewtimes \mb w\pp{j} \big)  - \frac 1 K \mb 1 \mb 1^T \big(\mb w\pp j^* \ewtimes \mb w\pp{j} \big) 
   \in \mathbb C^{K}.   
\eeq
As usual, the $k$-th row of $\mb E\laa \pp j $ is denoted by $ \mb e \laa \pp j[k]^T$,  
and the $k$-th element of $\mb e\lab \pp j $ by $ e \lab \pp j[k]$.
%denoted by 
%\beq
%\label{ela2}
%\mb e\lab \pp j = \big[  e \lab \pp j[1], \hdots,  e \laa \pp j[K] \big]^T \in \mathbb C^{K},
%\eeq
 Both  error terms occur at every iteration $j$ of the algorithm.
 We are now interested in the impact of such errors on $\mb w\pp{j}$. % (\ref{dlaw}).
 First of all, we notice that the $k$-th component of vector $\mb w\pp{j}$ in the ideal LA (\ref{law}) can be written as 
 \ifdouble
  \begin{align}
 w\pp j\ide[k] &\triangleq \frac K N \yr{k}^T \mb Y^H \mb v \pp j - (\mb v\pp j^H \mb R \mb v\pp j)  \cdot v\pp j[k] \nonumber \\ 
 & \ \ \ - \|\mb w\pp{j-1}\| \cdot v\pp{j-1}[k],
 \end{align}
 \else
 \beq
 w\pp j\ide[k] \triangleq \frac K N \yr{k}^T \mb Y^H \mb v \pp j - (\mb v\pp j^H \mb R \mb v\pp j)  \cdot v\pp j[k] - \|\mb w\pp{j-1}\| \cdot v\pp{j-1}[k],
 \eeq
 \fi
  by exploiting the expressions of $\alpha\pp j$ (\ref{laa}) and $\beta\pp j$ (\ref{lab}).
 We then define the error on $w\pp j[k]$ to be
\beq
\label{ewjk}
e_{w \pp j}[k] \triangleq w\pp j[k] -  w\pp j\ide[k].
\eeq
%where the terms $\mb v\pp j^H \mb R \mb v\pp j$ and $\|\mb w\pp{j-1}\|$ are the ``ideal'' values of $\alpha\pp j$ and $\beta\pp j$.
A closed-form expression of $e_{w(j)}[k]$ as a function of
 $\mb E\laa\pp j$ and $\mb e\lab \pp j $ on $e_{w(j)}[k]$ is given by the following Proposition. 
 Later, we provide an interpretation of this error expression, and we discuss its impact on the estimation of eigenvalues.

\begin{proposition}
\label{prop6}
Given a non-ideal AC scheme that introduces error terms $\mb E\laa\pp j$  and $\mb e\lab\pp j$ defined respectively in (\ref{ela1}) and (\ref{ela2}), the resulting error in $w\pp j[k]$ is given by
\ifdouble
\begin{align}
\label{prop6th1}
e_{w\pp j}[k ] &=  \frac 1 N (\mb e\laa\pp j[k])^H \yr{k}  - \frac {K^2} N \|  \mb e\laa\pp j[k] \|^2 \cdot v\pp j[k]  \nonumber \\
 & - \frac K N \left[ (\mb e\laa\pp j[k])^T \mb Y^H \mb v\pp M + \mb v\pp M^H \mb Y (\mb e\laa\pp j[k])^* \right]  v\pp j[k]  \nonumber  \\
&    -   \frac{K}{2 \|\mb w\pp{j-1}\|} e\lab\pp j[k]   \cdot v\pp{j-1}[k]   +  o(e\lab\pp j[k]) .
%\label{prop5th2}
\end{align}
\else
\begin{align}
\label{prop6th1}
e_{w(j)}[k] = & \frac 1 N (\mb e\laa\pp j[k])^H \yr{k} 
 -  \frac K N \left[ (\mb e\laa\pp j[k])^T \mb Y^H \mb v\pp M + \mb v\pp M^H \mb Y (\mb e\laa\pp j[k])^* \right] \cdot v\pp j[k]  \nonumber  \\
&   - \frac {K^2} N \|  \mb e\laa\pp j[k] \|^2 \cdot v\pp j[k]  -   \frac{K}{2 \|\mb w\pp{j-1}\|} e\lab\pp j[k]   \cdot v\pp{j-1}[k]   +  o(e\lab\pp j[k]) .
%\label{prop5th2}
\end{align}
\fi
%where $ (\mb e\laa[k])^T$ and $e\lab[k]$ are, respectively, the $k$-th row of $\mb E\laa$ and the $k$-th element of vector $\mb e\lab$.
\end{proposition}
\begin{IEEEproof}
The LA iteration   (\ref{law}), as well as the decentralized version (\ref{dlaw}), consists of three additive terms, therefore
the error (\ref{ewjk}) can be expressed as the sum of three separate terms:
\beq
e_{w \pp j}[k] = e^{\mathsf{(I)}}_{w \pp j}[k] + e^{\mathsf{(II)}}_{w \pp j}[k] + e^{\mathsf{(III)}}_{w \pp j}[k].
\eeq 
The first term originates from the computation of $\frac K N \diag \big\{  \mb Y \cdot \big[ \AC_N\big(\Diag (\mb v\pp j ^*) \cdot \mb Y\big) \big]^H  \big\}$ in (\ref{dlaw}), which is identical to the PM iteration. Thus, the first error term can be expressed using  (\ref{prop4j}) with $\mb E\pma\pp j$ replaced by $\mb E\laa\pp j$. The global error for all nodes is $\frac 1 N \diag\big[  \mb Y \big(\mb E\laa\pp j\big)^H \big]$, and its $k$-th element is
\beq\label{prop6proof1}
e^{\mathsf{(I)}}_{w\pp j}[k] = \frac 1 N (\mb e\laa\pp j[k])^H \yr{k}.\eeq

The second term arises from computation of $\alpha\pp j[k]$, which is done using the same vector $\mb z[k]$ obtained via AC in line \ref{dla1}. Therefore the error on $\alpha\pp j[k]$ depends again on $\mb E\laa\pp j$. Since $\alpha\pp j[k]$ is calculated as the square norm of  $\mb z[k]$ (line \ref{dlanorm}), the structure of the error is the same as that of $\enu[k]$ for the DPM (\ref{prop5th1}). By replacing $\mb E\pmb$ by $\mb E\laa\pp j$ in (\ref{prop5th1}) and multiplying by $v\pp j[k]$, we obtain the second part as 
\ifdouble
\begin{align}
e^{\mathsf{(II)}}_{w\pp j}[k] = &
\frac K N \left[ (\mb e\laa\pp j[k])^T \mb Y^H \mb v\pp M + \mb v\pp M^H \mb Y (\mb e\laa\pp j[k])^* \right] v\pp j[k] \nonumber  \\ 
&+ \frac {K^2} N  \|  \mb e\laa\pp j[k] \|^2 \cdot v\pp j[k].
\label{prop6proof2}
\end{align}
\else
\beq
\label{prop6proof2}
e^{\mathsf{(II)}}_{w\pp j}[k] =
\frac K N \left[ (\mb e\laa\pp j[k])^T \mb Y^H \mb v\pp M + \mb v\pp M^H \mb Y (\mb e\laa\pp j[k])^* \right] \cdot v\pp j[k]  + \frac {K^2} N \|  \mb e\laa\pp j[k] \|^2 \cdot v\pp j[k].
\eeq
\fi

The third part contains the error due to computation of $\beta\pp j[k]$, i.e., the norm of $\mb w\pp {j-1}$. 
The error arises from the AC algorithm used in line \ref{dla2} when computing $\|\mb w\pp {j-1}\|^2$; a nonlinearity is then introduced by the square root.
Using a first-order Taylor expansion (under the assumption that $e\lab\pp j [k]\ll \|\mb w\pp {j-1}\|$), we can write 
\begin{align}
\beta\pp j[k] &= \sqrt{ \|\mb w\pp {j-1}\|^2  + K  e\pp j\lab[k] } \label{p63start} \\
&=  \|\mb w\pp {j-1}\| \sqrt{          1 +    \frac{K}{\|\mb w\pp{j-1}\|^2 } e\lab\pp j[k]   }\\
&=   \|\mb w\pp {j-1}\| \left[          1 +    \frac{K}{2 \|\mb w\pp{j-1}\|^2} e\lab\pp j[k] + o(e\lab\pp j[k])   \right] \\
&=  \|\mb w\pp {j-1}\| +  \frac{K}{2 \|\mb w\pp{j-1}\|} e\lab\pp j[k] + o(e\lab\pp j[k]),   \label{p63end}
\end{align}
hence the third error term is given by
\beq
\label{prop6proof3}
e^{\mathsf{(III)}}_{w\pp j}[k] = 
 \frac{K}{2 \|\mb w\pp{j-1}\|} e\lab\pp j[k] \cdot v\pp{j-1}[k]    + o(e\lab\pp j[k]) .
\eeq
The resulting error (\ref{prop6th1}) is obtained by summation of  (\ref{prop6proof1}), (\ref{prop6proof2}), and (\ref{prop6proof3}).
\end{IEEEproof}

The error expressed  by Prop. \ref{prop6} is then propagated to the next iteration $j+1$ through another nonlinear step (line \ref{dlabk}). Using the same procedure of Eqs. (\ref{p63start})-(\ref{p63end}), the relation between $v\pp {j+1}[k]$ and $w\pp {j}[k]$ can be written as
\beq
\label{prop6coro}
v\pp {j+1}[k] = \frac{1}{1+    \frac{K}{\|\mb w\pp{j-1}\|^2 } e\lab\pp {j+1}[k]   + o(e\lab\pp {j+1}[k])    } \cdot  \frac {w\pp {j}[k]} {\|\mb w\pp {j}\|} .
\eeq
In summary, (\ref{prop6th1}) expresses the error on $\mb w\pp j$ given $\mb v\pp j$ and $\mb v\pp {j-1}$,
and   (\ref{prop6coro})  expresses the error on  $\mb v\pp {j+1}$ given $\mb w\pp j$. 
In principle, a closed-form error update rule could be derived from these expressions (as it was done for the DPM), but the resulting formula would be too complicated to provide additional insight.
We conclude the DLA analysis with two remarks.

\noindent
	\textit{1) } 
Following the same line of reasoning of \cite{paige76},   the error term in (\ref{prop6th1})  represents a \textit{loss of orthogonality} of the vectors $\mb v\pp j$, while (\ref{prop6coro}) results in a \textit{loss of unit norm}.
As documented in the literature, the loss of orthogonality leads to the appearance of so-called spurious eigenvalues at certain iterations (typically spurious eigenvalues are duplicates of existing eigenvalues already computed in the previous iterations). Some heuristics for the identification of spurious eigenvalues exist, such as the Cullum-Willoughby method \cite{cw}, which compares the eigenvalues of $\mb T$ (\ref{T}) with those of the same matrix without the first row and column. 
Alternatively, spurious values can be detected by comparing the eigenvalues of $\mb T$ at iterations $j$ and $j-1$.
In our application, another simple criterion can be derived by observing that the covariance matrix $\mb R$ has exactly $\min\{K,N\}$ non-zero eigenvalues, hence for iterations $j>\min\{K,N\}$ any new non-zero value is automatically identified as spurious.
We remark that all criteria for detection of spurious eigenvalues are compatible with the DLA, as they involve local post-processing of the output at individual nodes (matrices $\mb T[k]$ as defined in line \ref{dlaTk1} of Alg. \ref{dla}).

\noindent
\textit{2)}
The terms $\frac{K}{2 \|\mb w\pp{j-1}\|} e\lab\pp j[k]   \cdot v\pp{j-1}[k]  $ in (\ref{prop6th1})
 and $\frac{K}{\|\mb w\pp{j-1}\|^2 } e\lab\pp {j+1}[k]$  in (\ref{prop6coro})
 may be critical for the convergence of the algorithm in the event that $\|\mb w\pp{j}\| \approx 0$ (or, equivalently, $\beta\pp j \approx 0$) at some iteration $j$. 
 In the ideal LA, quoting \cite{golub},  ``a zero $\beta \pp j$ is a welcome event in that is signals the computation of an exact invariant subspace. However, an exact zero or even a small $\beta \pp j$ is a rarity in practice.''
This event is even more unlikely to happen in the case of the DLA: simulation results confirm that the values of $\beta\pp j$ are always far from zero. As a result, cases of divergence of the DLA were never observed.
 Nevertheless, the DLA turns out to be more sensitive to numerical problems than the DPM, as shown in Section \ref{simul}.

\subsection{Complexity}

\begin{table*}[tbp]
\centering
\small
\begin{tabular}{lcccc}
\hline
 & {Number of} &  {Number of}  &  {Number of information}  & {Number of required}\\ 
  & {$\AC_N$} &  {$\AC_1$}  &  { units exchanged per node}  &  {time periods}\\ 
\hline
\hline
\textbf{DPM} & $M+1$ & $1$ & $I(MN + N+ 1)d$ & $M+2$\\
\hline
\textbf{DLA} &  $M$ & $M$ & $I(MN + M)d$ & $2M$\\
\hline
\end{tabular}
\normalsize
\caption{Numerical complexity of DPM and DLA. Legend: $N=$ number of samples, $M = $ number of DPM/DLA iterations, $I =$ number of AC iterations, $d$ = node degree.}
\label{complex}
% \vspace{-1cm}
\end{table*}

For both the DPM and the DLA, complexity mainly arises from the repeated use of AC routines, resulting in communication overhead, time delays, and possible synchronization issues. 
We next compare the complexity of the two algorithms in terms of the following parameters: 
\textit{(i)} number of calls to functions {$\AC_N$}  and {$\AC_1$};
\textit{(ii)} total number of ``information units'' exchanged over the wireless channel by one node with degree (number of neighbors) $d$, where an information unit is defined as the number of bits used to encode a scalar; 
\textit{(iii)}  number of algorithmic steps,   defined as individual tasks that must be performed in a sequential way due to input-output dependency (i.e., step $n$ cannot start until step $n-1$ is completed).

The above performance figures are reported in Table \ref{complex} for the two algorithms. 
For simplicity it is assumed that all instances of AC have the same number of iterations, $I$.
It can be observed that the total number of calls to consensus routines is slightly higher for the DLA (assuming $M>2)$: the DPM requires one vector-AC per iteration (line \ref{dpm1}) in addition to another vector-AC and a scalar-AC for eigenvalue computation (lines \ref{dpm2}-\ref{dpm3}), whereas the DLA uses one vector-AC and one scalar-AC in each iteration (lines \ref{dla1}-\ref{dla2}).
The amount of information sent over the air is nearly the same, as shown in the third column of the table.
However, the DLA is significantly more complex than the DPM  in terms of the number of algorithmic steps: in the case of the DLA, each iteration consists of two sequential steps (vector computation and normalization) that cannot be parallelized, hence the total number of steps for the DLA is nearly twice that of the DPM. 

The above results suggest that the values of $M$, $N$, and $I$ should be kept as  small as possible in order to reduce complexity and improve the reactivity of the algorithms, especially when used in real-time detection applications. 
In particular, the number of samples can be very small  if the number of nodes $K$ is large enough
(large networks are the natural scenario of application for decentralized algorithms). 
For example, in a network of $40$ nodes, $10$ samples per node are sufficient to detect a signal with SNR of $7$dB with high probability (see Fig. \ref{roc10} in Section \ref{simul}).

We now briefly analyze the computational complexity for individual nodes. 
For the DPM, the dominant factor is the vector product of  line \ref{dpmloc1}. Since the vector size is $N$ and the product is computed at every iteration, the   computational complexity per node scales as $O(MN)$.
In the DLA, every iteration involves computing the norm of a vector of size $N$ (line \ref{dlanorm}) and a vector product of the same size (line \ref{dlawk}), hence the computational complexity per node scales as $O(MN)$ as well. In addition, the DLA involves local calculation of the eigenvalues of the tridiagonal matrix $\mb T[k]$ (line \ref{dlaTk}), which has complexity $O(M^2)$. The dominant term between $O(MN)$ and $O(M^2)$ depends on the relative values of $M$   and $N$.

\section{Application to Spectrum Sensing}
\label{app}
\label{cr}

%In this section we illustrate two applications of the proposed decentralized eigenvalue algorithms. 
%The first application is eigenvalue-based spectrum sensing in cognitive radio networks \cite{blindly, penna_cl,wei_ofdm,zeng_tcom,zeng_glrt,glrd,bianchi,nadler_icc,wei}. In this scenario, multiple sensor nodes cooperate to detect the presence of a primary signal in a given frequency band. A decision about signal presence or absence is made upon receiving $N$ signal samples at each sensor. The main challenge is to achieve reliable  detection of weak signals (low SNR) with a limited number of samples.
%The second application is estimation of the number of signal sources in decentralized sensor networks or virtual MIMO systems. This problem can be addressed, again, by applying estimation techniques based on the eigenvalues of the received sample covariance matrix \cite{viberg,wax,itc_perf,itc_perf2,nadler_tsp}.
%The estimation of the number of signals in a multi-sensor model may be sometimes  the first step of more sophisticated applications, such as estimation of the direction of arrival (DoA) of unknown signals \cite{music} or anomaly detection via principal component analysis \cite{huang}.

%\subsection{Spectrum Sensing in Cognitive Radio Networks}

\begin{table*}[tbp]
\centering
\small
\begin{tabular}{cccc}
\hline
  \textbf{{Name}} &  \textbf{{Test statistic}}  &  \textbf{{Application scenario}} & \textbf{{Ref.}}\\ 
\hline
\hline
Roy's  test (RT) & $\TR \triangleq \lambda_1 / \sigma^2$  & $P=1$, known $\sigma^2$ & \cite{nadler_icc,blindly,wei_ofdm} \\
\hline
GLR test (GT) &  $\TG \triangleq \lambda_1/\sum_{i=1}^K \lambda_i$ & $P=1$, unknown $\sigma^2$ & \cite{zeng_glrt,glrd,bianchi} \\
\hline
Sphericity test (ST) &  $\TS \triangleq {\prod_{i=1}^K \lambda_i} \big/ {\left( \frac 1 K \sum_{i=1}^K \lambda_i \right)^K} $   & $P>1$, finite SNR & \cite{zeng_glrt,glrd,wei} \\
\hline
John's  test (JT) &  $ \TJ \triangleq {\sum_{i=1}^K \lambda^2_i} \big/ {\left(  \sum_{i=1}^K \lambda_i \right)^2}$ & $P>1$, low SNR  & \cite{wei} \\
\hline
\end{tabular}
\normalsize
\caption{Eigenvalue-based tests for multi-sensor signal detection.}
\label{tests}
% \vspace{-1cm}
\end{table*}

We consider a distributed  spectrum sensing scenario, where multiple sensor nodes cooperate to detect the presence of a primary signal in a given frequency band. A decision about signal presence or absence is made upon receiving $N$ signal samples at each sensor. The main challenge for cognitive networks is to achieve reliable signal detection with a limited number of samples.
Mathematically, the 
problem is a binary hypothesis test between a ``signal-plus-noise hypothesis'' ($\mc H_1$) and a ``noise-only'' hypothesis ($\mc H_0$) based on the received samples $\mb Y$.
The network is assumed to operate under homogeneous conditions, i.e., the same hypothesis ($\mc H_0$ or $\mc H_1$) holds for all sensors during the entire sensing period.
Given the model already introduced in Section \ref{ps} and assuming zero-mean complex Gaussian noise, the signal vector received at a given time instant $n$ at the $K$ sensors can be written under $\mathcal H_0$  as
\begin{equation}
\mb{y}(n)|_{\mathcal{H}_0} = \bm{\eta}(n),
\end{equation}
where $\bm{\eta}(n) \sim \mc{N}_{\mathbb{C}} (\mb{0}, \sigma^2 \mb{I})$. 
Under
$\mathcal H_1$, the received vector is
\begin{equation}
\label{yH1}
\mb{y}(n)|_{\mathcal{H}_1} = %\mb {x}(n) + \bm \eta(n) = 
\mb{H} \mb s(n) + \bm{\eta}(n),
\end{equation}
where $\mb s(n)  \sim \mc{N}_{\mathbb{C}} (\mb{0}, \Diag([\sigma^2_1, \hdots, \sigma^2_P])) \in \mathbb C^P$ is a vector of  signal samples transmitted by $P$  sources (primary users), modeled as zero-mean Gaussian mutually uncorrelated random variables, 
and  $\mb H = [\mb{h}_1, \hdots, \mb h_P] \in \mathbb C^{K \times P}$ is a complex channel matrix whose columns represent the channels between the $P$ signal sources  and the $K$
sensors. The channel coefficients are assumed to be
%, where each element $h_{k}$ is the channel coefficient referred to receiver $k$ (for simplicity,
unknown but constant during the sensing period\footnote{Assuming the sensing period to be shorter than the coherence time of the channel  is reasonable in multi-sensor spectrum sensing applications, where accurate detection is achieved already at low sample size.
%Also note that an opposite setting, i.e., constant-modulus signals transmitted over a Rayleigh fading channel, results in an identical mathematical model for $\mb x(n)$.
}. 
The  SNR for the $i$-th signal source  is  defined (under  $\mc H_1$) as
$\rho_i \triangleq 
%\frac{\mathbb{E} \; \| \mb{x}(n) \| ^2 } {\mathbb{E} \; \| \bm{\eta}(n) \| ^2} 
 \| \mb{h}_i \|^2 \sigma^2_i  / \sigma^2$.
%which can be interpreted as the average of the local SNRs at the $K$ sensors. 
Let $T(\mb Y)$ be a generic test statistic for signal detection, computed from the signal samples, and let $\th$ be the associated decision threshold, such that the detector decides for $\mc H_1$ if  $T(\mb Y) > \th$ and for $\mc H_0$ otherwise. Then, false-alarm and detection probabilities are defined, respectively, as
$
\Pfa \triangleq \Pr[T(\mb Y) > \th | \mc H_0] 
$
and
$
\Pd \triangleq \Pr[T(\mb Y) > \th | \mc H_1]$.
%Since $\Pfa$ can be in certain cases calculated analytically and does not depend on the (usually unknown) SNR, 
Detectors are typically calibrated so as to achieve a fixed false-alarm rate $\Pfa = \alpha$, i.e., the threshold is set as $\th(\alpha)$.

Signal detection in the above-described setting has been extensively studied in the cognitive radio literature, e.g., \cite{blindly,penna_cl,wei_ofdm,zeng_tcom,zeng_glrt,glrd,bianchi,nadler_icc,wei}, leading to the derivation of several possible test statistics  $T(\mb Y)$.
Most of such statistics are functions of the eigenvalues of the sample covariance matrix (some examples are reported in Table \ref{tests}) and, 
as such, they can be computed in a decentralized network   by applying the DPM or the DLA proposed in this paper. 
%A number of eigenvalue-based tests characterized by some optimality properties are summarized in Table \ref{tests}. 
%For all mentioned tests, optimality is meant under the common assumptions of Gaussian signal, Gaussian noise, and asymptotical sample size $N$.
%Specifically, the first detector, referred to as ``Roy's largest root test'' (RT),  is the optimal test statistic in the Neyman-Pearson sense under single-signal  ($P=1$) and finite SNR assumption (see \cite{nadler_icc} and references therein); it was introduced in the cognitive radio literature with different derivations in \cite{blindly} and \cite{wei_ofdm}.
%Again for $P=1$, but with unknown noise variance, an optimal detector according to the generalized likelihood ratio (GLR) principle is given by the largest eigenvalue scaled by the trace of $\mb R$ (see \cite{zeng_glrt,glrd,bianchi,nadler_icc}). We refer to this detector as GLR test, or GT for short. 
%In the case of multiple signal sources ($P>1$), the optimal detector for finite SNR is given by the so-called sphericity test (ST) \cite{zeng_glrt,glrd,wei}, while a locally best invariant procedure, suitable for low SNR, is given by a method known as ``John's test'' (JT) \cite{wei}. 
%Definitions of the aforementioned test statistics are reported in the table.
% \ref{tests}.
%All the above detectors can be reformulated in a decentralized manner by using the DPM or the DLA. 
The problem reduces to computing, for each sensor $k$, a local version of the adopted test statistic, say   $\hat T_i[k]$ with $i \in \{\mathsf{R}, \mathsf{G}, \mathsf{S}, \mathsf{J}\}$. This is obtained directly as a function of the local eigenvalue estimates $\hat \lambda_j[k]$ (with $j=1$ for the DPM, which is sufficient to compute the RT statistic; and $1\leq j \leq K$ for the DLA).
Then, each node tests the local statistic $\hat T_i[k]$ against the predefined threshold $\th(\alpha)$ and decides for $\mc H_0$ or $\mc H_1$ according to the rule
\beq
\label{localtest}
\hat T_i[k] \maggmin \th(\alpha).
\eeq
In order to average out the numerical errors introduced by non-ideal AC at different nodes\footnote{Numerical errors may result in the problem of different decisions at different nodes if $\hat T_i[k]$ is very close to $\th(\alpha)$.}, a final round of AC can be executed taking as inputs the local statistics $\hat T_i[k]$. The result
\beq
\hat T'_i[k] = \AC^t_1[k]\big( \hat T_i[k] \big)
\eeq 
shall be used in (\ref{localtest}) instead of $\hat T_i[k]$.  

The popular cooperative energy detector, i.e., the test based on the (possibly weighted) sum of the received signal energies at different sensors, also admits a natural decentralized implementation via AC algorithms.
This problem was investigated in \cite{ed_cons}. 
Decentralized energy detection is computationally simpler than eigenvalue-based techniques, but clearly inherits the well-known shortcomings of energy detection (suboptimality in multi-sensor settings and sensitivity to noise uncertainty). 

In some applications, the goal is not only to discriminate between $\mc H_0$ and $\mc H_1$, but to estimate the number of signals $P$.
This problem can be addressed again by eigenvalue-based estimators, such as the well-known Wax and Kailath's information-theoretic criteria \cite{wax}, or the recent  random matrix theory (RMT) estimator proposed in \cite{nadler_tsp}. In all cases, the estimated number of signals is estimated as 
\beq
\hat P = \arg\min_{0\leq q\leq K} T(q; \lambda_1, \hdots, \lambda_K ),
\eeq
where $T(\cdot)$ is a  function of the eigenvalues and, once again, can be computed locally from the DLA estimates $ \hat \lambda_1, \hdots, \hat \lambda_K$.

 \section{Simulation Results}
 \label{simul}

  \begin{figure}
	\centering
		\includegraphics[width=\wi]{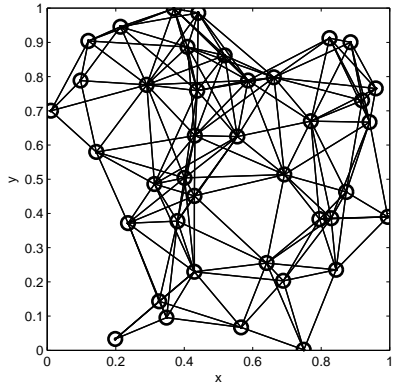}
	\caption{Network topology.}
		\label{topo}
\end{figure} 
 
   \begin{figure}
	\centering
		\includegraphics[width=\wi]{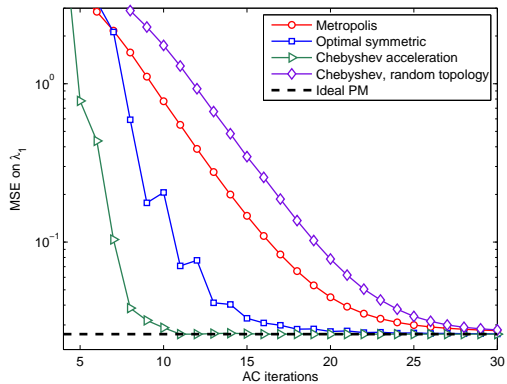}
	\caption{Comparison of different AC algorithms.}
		\label{compare_ac}
\end{figure} 
 
 For the purpose of numerical evaluation of the proposed algorithms, we consider a randomly generated network consisting of $K=40$ nodes, as depicted in Fig. \ref{topo}. Edges in the graph represent communication links between pairs of nodes. The received signal samples $\mb Y$ are randomly generated at every Monte Carlo iteration according to the signal-plus-noise model described in Section \ref{cr} (case $\mc H_1$), with the following parameters: $N=10$ samples per node, noise variance $\sigma^2 = 1$, one Gaussian signal source ($P=1$) with SNR $\rho = 5$dB.
 
 We first turn our attention to the convergence of the largest eigenvalue, expressed in terms of the mean square error (MSE) of $\hat \lambda_1$ with respect to $\lambda_1$.
 In practice, the MSE is calculated as an average over $3000$ Monte Carlo simulations. In Fig. \ref{compare_ac} we compare the performance of different AC algorithms when applied in the DPM (Alg. \ref{dpm}). The number of DPM iterations is fixed to $M=20$, while the number of AC iterations ($I$) varies from $5$ to $30$ (x-axis in the plot). 
 We consider four alternative AC schemes, namely two versions of traditional AC \cite{boyd} -- with Metropolis weights (a heuristic based on the graph Laplacian) and with optimal symmetric weights (resulting from the solution of a convex optimization problem) -- and two versions of AC with Chebyshev acceleration \cite{renato_cheb} --  assuming first a fixed network topology and then one with $3\%$ link failure probability. The lowest achievable bound is given by the ideal PM (i.e., a DPM with exact AC).
 The Chebyshev acceleration algorithm turns out to significantly outperform traditional AC even with optimized weights (under deterministic settings). For this reason we adopt Chebyshev-accelerated AC as the standard consensus  scheme in all the following simulations.

    \begin{figure}
	\centering
		\includegraphics[width=\wi]{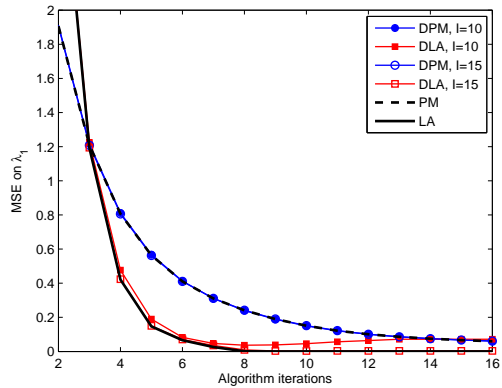}
	\caption{Convergence of $\lambda_1$.}
		\label{conv_l1}
\end{figure}

In Fig. \ref{conv_l1} we compare the DPM against the DLA. We observe that the DLA exhibits a faster convergence rate, but is more sensitive than the DPM to numerical errors introduced by imperfect consensus:  the DLA error  converges  to $0$ for $I=15$ AC iterations, but not for $I=10$, whereas the performance of the DPM is practically identical to that of the ideal PM already for $I=10$ (and even for smaller values). In other words, the performance of the DLA compared to the DPM is better in absolute terms (especially with few algorithm iterations), but worse in relative terms. This behavior is consistent with the analysis presented in Section \ref{perf} and can be understood intuitively by noticing that each step of the DLA uses  AC twice, in contrast to the DPM where AC is used only once per iteration. 

\begin{figure}
	\centering
	\subfigure[$I=20$]{
		\includegraphics[width=\wi]{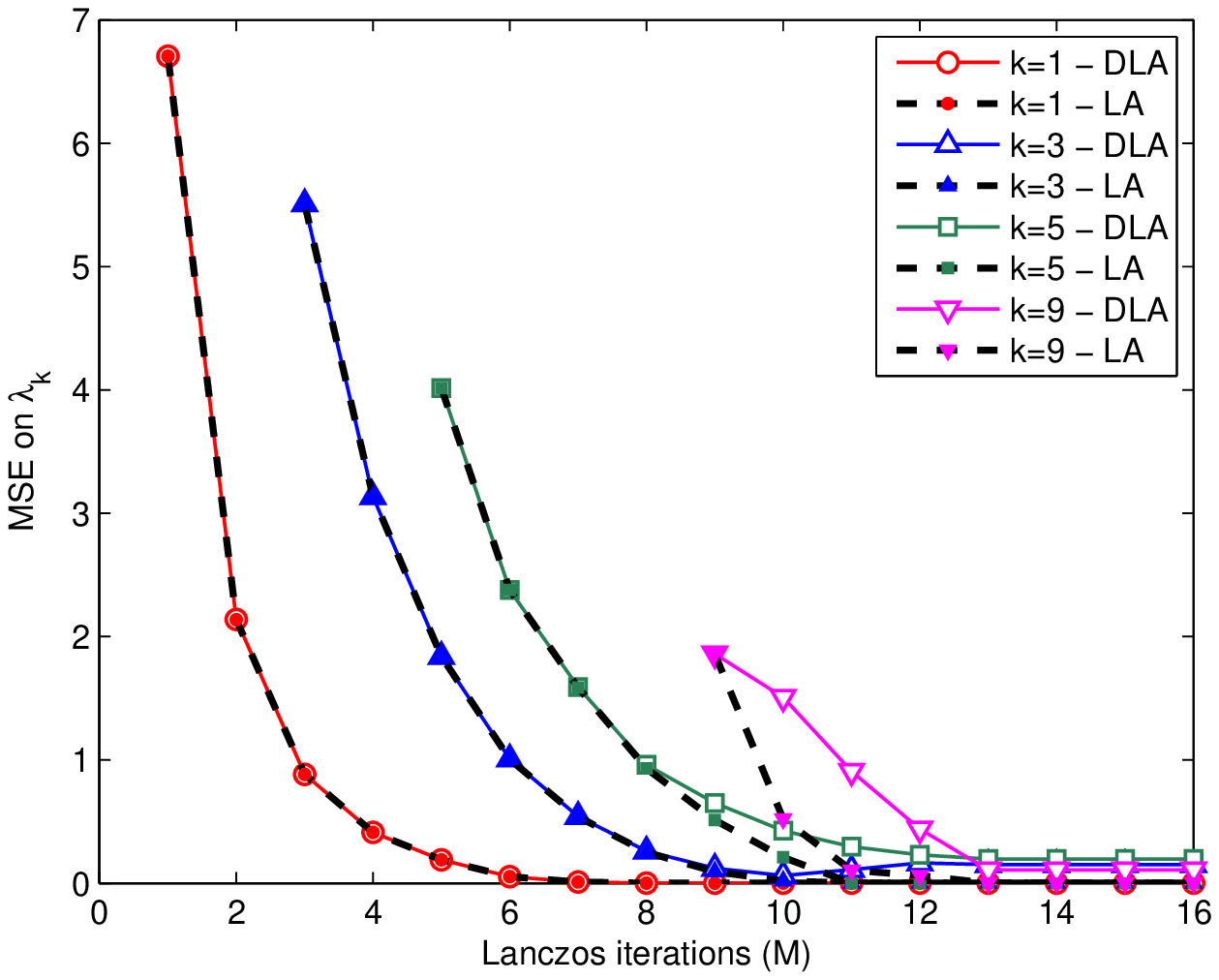}
	\label{convmult20}}
		\subfigure[$I=30$]{
		\includegraphics[width=\wi]{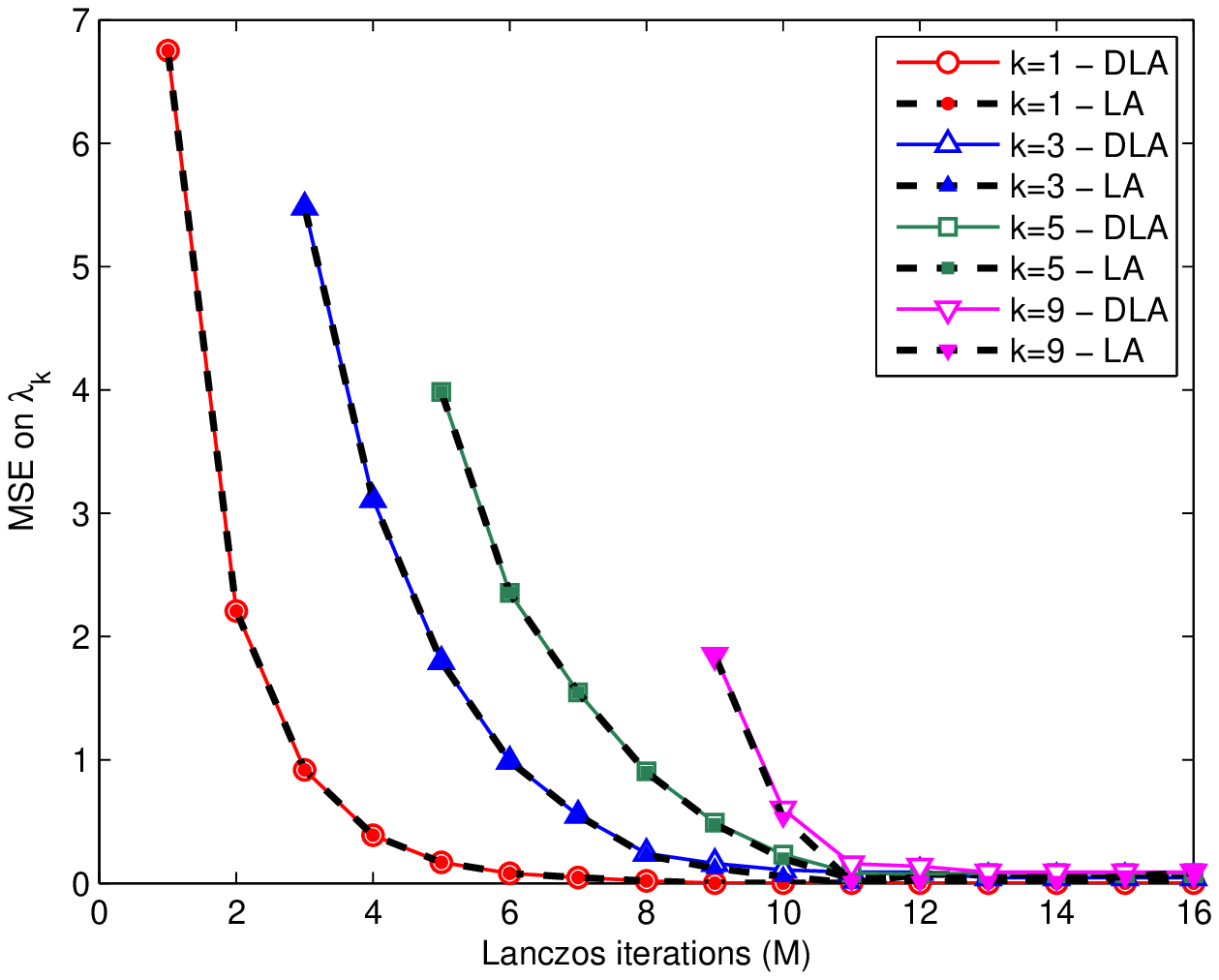}
	\label{convmult30}}
%		\subfigure[$I=40$]{
%		\includegraphics[width=\wi]{Figs/convmult40}
%	\label{convmult40}}
	\caption{Convergence of multiple eigenvalues.}
	\label{convmult}
	\end{figure}

 We next evaluate the performance of the DLA when estimating multiple eigenvalues. The results, shown in Fig. \ref{convmult}, indicate that the higher is the order of eigenvalues to be estimated, the more iterations are needed.
 The reason is that, by definition of the LA, the $j$-th eigenvalue cannot be estimated until the $j$-th iteration; in addition, numerical errors sometimes cause the appearance of spurious eigenvalues (see Section \ref{perf_dla}) which, even when correctly identified, introduce a one-step delay in the algorithm. 
 One way to mitigate the numerical problems that affect the estimation of high-order eigenvalues is to improve the precision of the AC routine by tuning the parameter $I$. For example, according to our simulations,  $20$ AC iterations are sufficient to achieve an accurate estimation of the $\lambda_1$ and $\lambda_3$ (Fig. \ref{convmult20}), whereas $30$ iterations are necessary for $\lambda_5$ and $\lambda_9$ (Fig. \ref{convmult30}).

\begin{figure}
	\centering
	\subfigure[$M=5$]{
		\includegraphics[width=\wi]{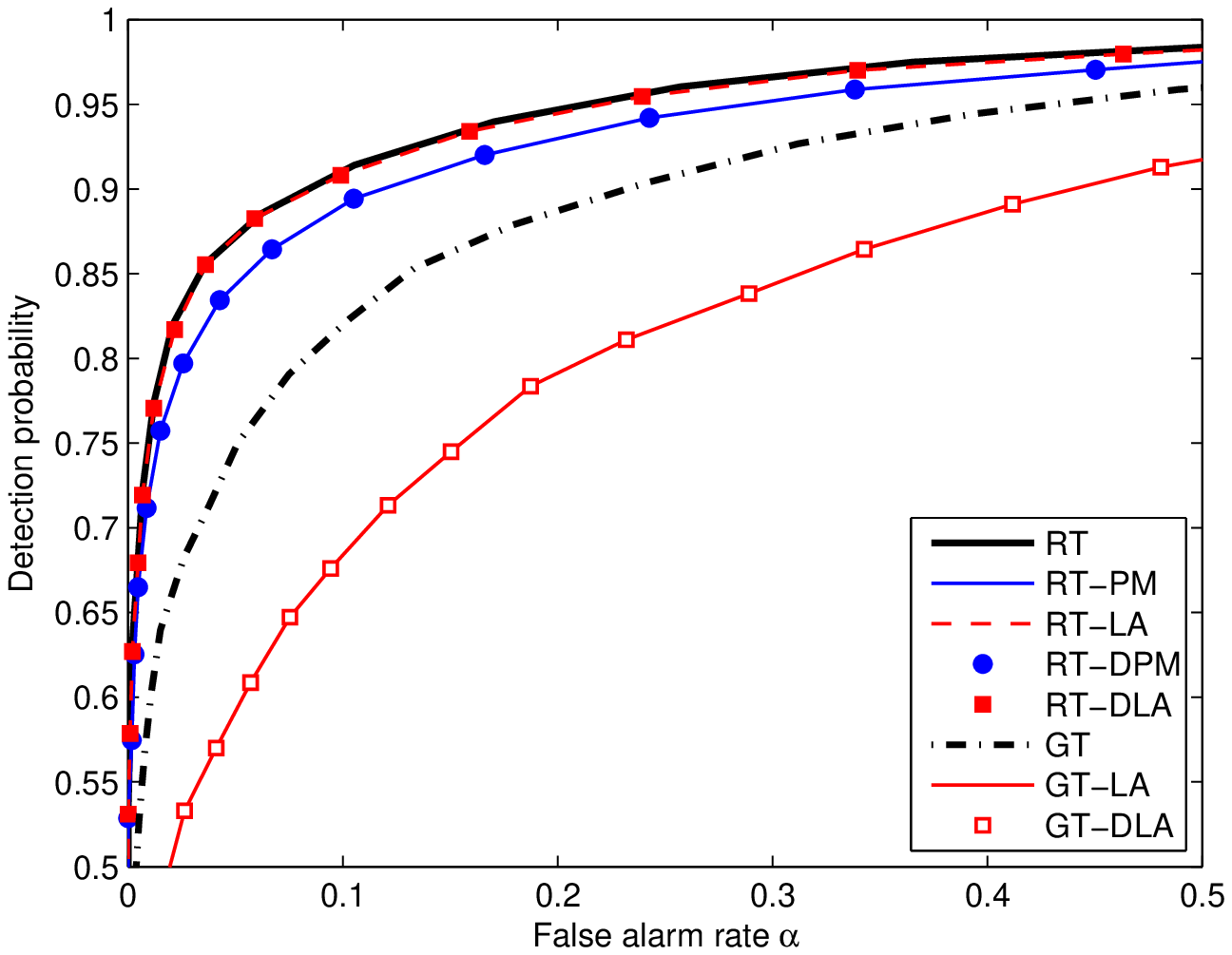}
	\label{roc5}}
		\subfigure[$M=10$]{
		\includegraphics[width=\wi]{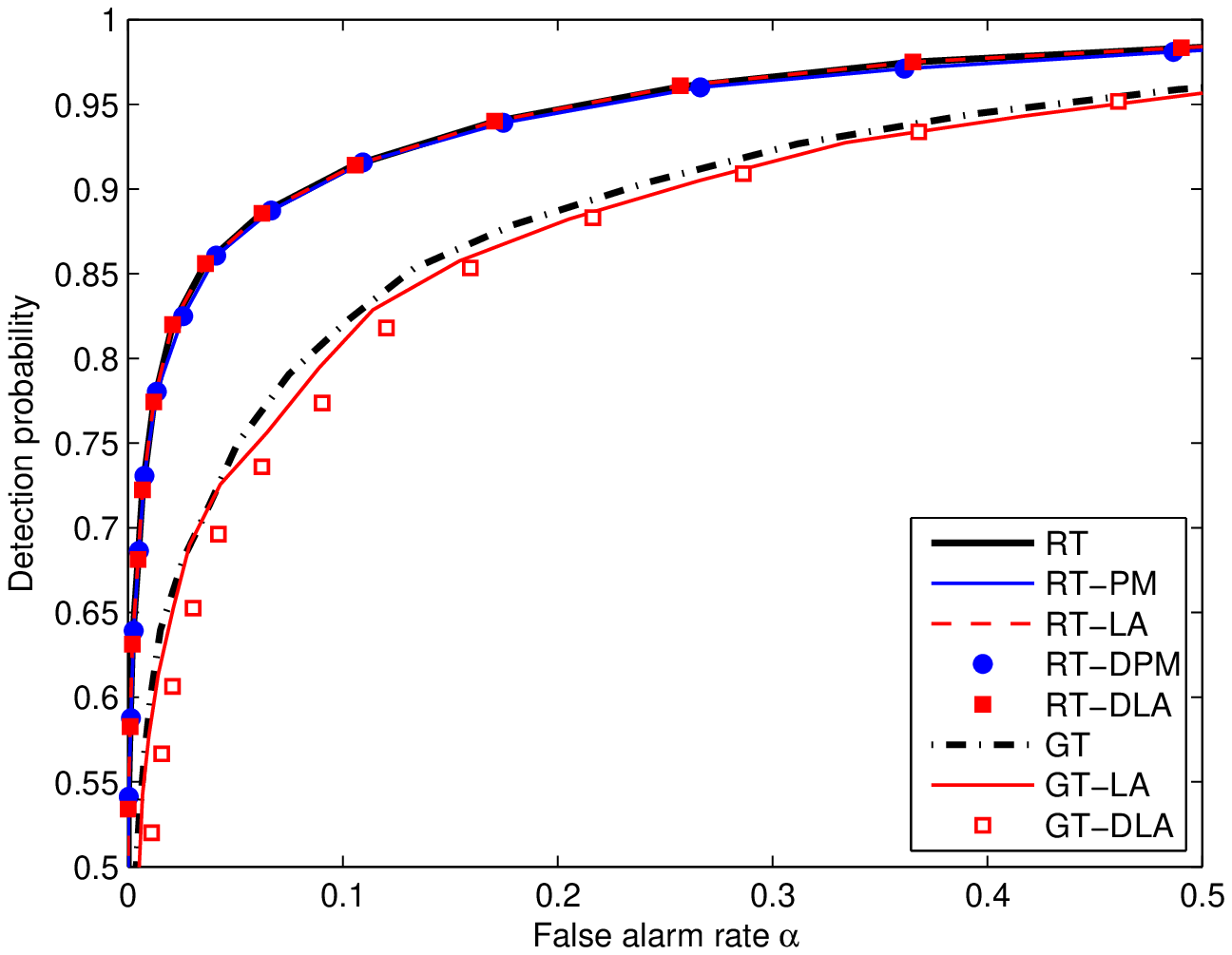}
	\label{roc10}}
	\caption{Signal detection: ROC curves.}
	\label{roc}
	\end{figure}

 We now illustrate  an application of the proposed DPM and DLA for spectrum sensing in a distributed cognitive radio network. We assume again the same network topology of Fig. \ref{topo}, with $K=40$ and $N=10$. 
 According to the model introduced in Section \ref{cr}, the samples under $\mc H_0$ are Gaussian distributed with variance $\sigma^2=1$, while under $\mc H_1$ we have one Gaussian signal component ($P=1$) with SNR $\rho=7$dB. We test the performance of two signal detectors: the RT and the GT, defined respectively as $\TR$ and $\TG$ in Tab. \ref{tests}. The RT is a test of significance of the largest eigenvalue ($\lambda_1$) alone, hence it can be implemented in a decentralized setting using either the DPM or the DLA.  The GT, in contrast, involves all eigenvalues and requires the use of the DLA. 
 In Fig. \ref{roc} we compare: \textit{(i)} the ideal performance of  RT and GT using the  eigenvalues of $\mb R$ computed exactly by a fusion center with perfect communication links; 
 \textit{(ii)} the performance achieved when the eigenvalues are computed by the PM or LA (assuming ideal AC), respectively for $M=5$ and $M=10$; 
 and \textit{(iii)} the performance achieved when the eigenvalues are computed by the DPM or DLA, with $I=30$ iterations.
 The results show that, after $M=5$ algorithm iterations (Fig. \ref{roc5}), the RT using LA or DLA already attains nearly-ideal performance, while the RT with DPM is slightly suboptimal; for the GT, on the other hand, $5$  iterations are not enough to reach convergence of all eigenvalues (note that the performance gap is due to the LA itself and not to the decentralized version). After $M=10$ iterations (Fig. \ref{roc10}),   all versions of the RT converge to the ideal RT bound, and the gap of LA- and DLA-GT compared to the ideal GT is dramatically reduced.

%	   \begin{figure}
%	\centering
%		\includegraphics[width=\wi]{Figs/nsig_temp}
%	\caption{Multiple signal detection.}
%		\label{nsig}
%\end{figure} 
%% 
% 
% 
%
 \section{Conclusion}
 \label{concl}
 
 In this paper we have proposed and analyzed two general-purpose algorithms that can be used for computing sample covariance eigenvalues in distributed wireless networks.
 As an application, we have considered spectrum sensing in cognitive networks, and we have shown that numerous eigenvalue-based tests for single-signal and multiple-signal detection can be implemented in a decentralized setting by using the proposed algorithms. 
 Such decentralized signal detection techniques enable sensor nodes to compute global test statistics locally,  thereby performing hypothesis tests without relying on a fusion center. Decentralized approaches also provide additional robustness to node failures or Byzantine attacks. 
 
The two proposed algorithms -- the DPM and the DLA -- have different strengths and drawbacks. The DPM is less complex, more robust to numerical problems, and provably convergent even in the presence of non-ideal AC (under the conditions of Proposition \ref{convdpm}); however, it can estimate only the largest eigenvalue, and under ideal assumptions its convergence rate is slower than that of the DLA. On the other hand, the DLA is able to estimate all eigenvalues (albeit with increasing complexity with the number of eigenvalues) and, even in the presence of AC errors,  provides faster initial convergence than the DPM; however, it is more complex and more sensitive to numerical errors, and requires some post-processing in order to remove ``spurious'' eigenvalues. 

Evaluated through simulations, both algorithms exhibit good performance in practical conditions (non-ideal AC algorithms, small number of samples) already after few iterations. In particular, convergence is very fast in the case of the largest eigenvalue, which results in high-performing distributed signal detectors based on the largest eigenvalue (referred to as RT-DPM and RT-DLA in Fig. \ref{roc}).
Other possible fields of applications of the proposed algorithms are distributed anomaly detection in wireless sensor networks and signal feature estimation in distributed antenna arrays.

\section*{Acknowledgments}
 \label{ackno}
The authors are grateful to colleagues Jafar Mohammadi and Meng Zheng for their valuable comments and discussions on some of the topics of this paper, and to Renato L. G. Cavalcante for providing the simulation code from \cite{renato_cheb}.

\end{document}